\documentclass[twocolumn]{aastex631}
\usepackage{pgfplots}
\usepackage{pgfplotstable}
\usepackage{color}
\usepackage{enumerate}
\usepackage{amsmath}
\usepackage{tikz}
\usetikzlibrary{shapes.geometric, arrows, fit}
\usetikzlibrary{calc} 
\usepackage{appendix}
\revised{September 1, 2022}
\received{\today}
\accepted{??? xx, 2022}

\submitjournal{ApJ}

\newcommand{\nuebar}{\bar{\nu}_e}
\newcommand{\nue}{\nu_e}

\newcommand{\nuxbar}{\bar{\nu}_x}

\renewcommand{\sec}{\mathrm{s}}

\defcitealias{Suwa_2019}{Paper~I}
\defcitealias{Nakazato_2022}{Paper~II}

\shorttitle{Observing SN Neutrinos. VI. Realistic Background}
\shortauthors{Nakanishi et al.}
\graphicspath{{./}{figures/}}

\begin{document}

\title{Observing Supernova Neutrino Light Curves with Super-Kamiokande.\\
VI. A Practical Data Analysis Technique Considering Realistic Experimental Backgrounds}

\author[0000-0003-4408-6929]{Fumi Nakanishi}
\affiliation{Department of Physics, Okayama University, Okayama 700-8530, Japan}

\author[0000-0001-6330-1685]{Ken'ichiro Nakazato}
\affiliation{Faculty of Arts and Science, Kyushu University, Fukuoka 819-0395, Japan}

\author[0000-0003-3273-946X]{Masayuki Harada}
\affiliation{Kamioka Observatory, Institute for Cosmic Ray Research, The University of Tokyo, Gifu 506-1205, Japan}

\author[0000-0003-0437-8505]{Yusuke Koshio}
\affiliation{Department of Physics, Okayama University, Okayama 700-8530, Japan}
\affiliation{Kavli Institute for the Physics and Mathematics of the Universe (Kavli IPMU, WPI), Todai Institutes for Advanced Study, The University of Tokyo, Kashiwa, Chiba 277-8583, Japan}

\author[0000-0002-9234-813X]{Ryuichiro Akaho}
\affiliation{Faculty of Science and Engineering, Waseda University, Tokyo 169-8555, Japan}

\author[0000-0003-4136-2086]{Yosuke Ashida}
\affiliation{Department of Physics, Tohoku University, Sendai, Miyagi 980-8578, Japan}

\author[0000-0003-1409-0695]{Akira Harada}
\affiliation{National Institute of Technology, Ibaraki College, Hitachinaka, Ibaraki 312-8508, Japan}
\affiliation{Interdisciplinary Theoretical and Mathematical Sciences Program (iTHEMS), RIKEN, Wako, Saitama 351-0198, Japan}

\author[0000-0002-0827-9152]{Masamitsu Mori}
\affiliation{Division of Science, National Astronomical Observatory of Japan, 2-21-1 Osawa, Mitaka, Tokyo 181-8588, Japan}
\affiliation{National Institute of Technology, Numazu College, Numazu, Shizuoka 410-8501, Japan}

\author[0000-0002-9224-9449]{Kohsuke Sumiyoshi}
\affiliation{National Institute of Technology, Numazu College, Numazu, Shizuoka 410-8501, Japan}

\author[0000-0002-7443-2215]{Yudai Suwa}
\affiliation{Department of Earth Science and Astronomy, The University of Tokyo, Tokyo 153-8902, Japan}
\affiliation{Center for Gravitational Physics and Quantum Information, Yukawa Institute for Theoretical Physics, Kyoto University, Kyoto 606-8502, Japan}

\author[0000-0002-0969-4681]{Roger A. Wendell}
\affiliation{Department of Physics, Kyoto University, Kyoto 606-8502, Japan}
\affiliation{Kavli Institute for the Physics and Mathematics of the Universe (Kavli IPMU, WPI), Todai Institutes for Advanced Study, The University of Tokyo, Kashiwa, Chiba 277-8583, Japan}

\author[0000-0001-7305-1683]{Masamichi Zaizen}
\affiliation{Department of Earth Science and Astronomy, The University of Tokyo, Tokyo 153-8902, Japan}



\begin{abstract}
    Neutrinos from supernovae, especially those emitted during the late phase of core collapse, are essential for understanding the final stages of massive star evolution. 
    We have been dedicated to developing methods for the analysis of neutrinos emitted during the late phase and observed at Super-Kamiokande (SK). 
    Our previous studies have successfully demonstrated the potential of various analysis methods in extracting essential physical properties; however, the lack of background consideration has limited their practical application.
    In this study, we address this issue by incorporating a realistic treatment of the experimental signal and background events with the on-going SK experiment.
    We therefore optimize our analysis framework to reflect realistic observational conditions, including both signal and background events.
    Using this framework we study several long-time supernova models, simulating the late phase neutrino observation in SK and focusing in particular on the identification of the last observed event.
    We discuss the possibility of model discrimination methods using timing information from this last observed event.
\end{abstract}


\section{Introduction}
\label{cap:intro}
Stars exceeding $8M_{\odot}$ undergo core-collapse supernovae (CCSNe) when their internal pressure can no longer support their own mass.
Throughout the explosion neutrinos are released in various phases, taking 99\% of the gravitational binding energy of the star with them. 
This makes neutrinos a unique probe of stellar core physics~\citep[see, e.g.,][]{1987STIN...8729393B,2006RPPh...69..971K,2012ARNPS..62...81S,2017hsn..book.1575J,2018JPhG...45d3002H,2018MNRAS.475L..91T,2021MNRAS.500..696N}.
and makes their observation a key target of astrophysics research.
The neutrino emission process can be separated into several stages, including the pre-supernova~\citep{2020ARNPS..70..121K} phase, the neutronization burst~\citep{2003ApJ...592..434T}, the accretion phase~\citep{2013ApJ...762..126O}, and the cooling phase~\citep{2014PASJ...66L...1S,Nakazato_2018}.
During this cooling phase a proto-neutron star (PNS) is formed at the center of the supernova which eventually cools via neutrino emission over tens of seconds to become a cold neutron star.
The number and spectra of neutrinos emitted during this phase depend primarily on the mass and radius of the PNS, making their light curve less uncertain than those emitted during earlier phases.
For this reason, long-term CCSN simulations that include neutrino emission during the cooling phase have recently garnered attention in the field~\citep{Nakazato_2013,Suwa_2019,2021PTEP.2021b3E01M,2021PhRvD.103b3016L}. 

On the observation side supernova neutrinos have only been observed once:
During the explosion of SN-1987A in 1987 a total of 24 neutrino events were detected in the Kamiokande~\citep{1987PhRvL..58.1490H}, IMB~\citep{1987PhRvL..58.1494B}, and Baksan~\citep{1988PhLB..205..209A} detectors. 
Despite limited statistics this observation was nonetheless able to provide positive support for the neutrino-heating scenario that had recently been proposed to achieve successful explosion in simulations~\citep{1987PhLB..196..267S, 1989ApJ...340..426L}. 
Nevertheless, given the comparatively small detector sizes and the roughly 50~kpc distance to SN~1987A the data were insufficient to construct a detailed picture of the explosion mechanism. 

Current- and next-generation detectors will provide a larger, richer data set from the next CCSN.
Indeed, several large water Cherenkov and large liquid scintillator detectors, such as Super-Kamiokande~\citep[SK;][]{2003NIMPA.501..418F}, KamLand~\citep{2022ApJ...934...85A}, Hyper-Kamiokande~\citep[HK;][]{2018arXiv180504163H}, JUNO~\citep{2016JPhG...43c0401A}, and DUNE~\citep{2020JInst..15.8008A} are in operation or under construction and each has supernova neutrino observation as one of its science targets.
These are highly sensitive detectors capable of determining the arrival time and energy of individual neutrinos.
In particular, the currently operating Super-Kamiokande detector is expected to observe 1000 to 10,000 neutrino events if a supernova explosion occurs in our galaxy, which will provide valuable information on the explosion mechanism and progenitors. 
In addition, SK has the ability to determine the direction of the supernova with an accuracy of 3 to 7 degrees~\citep{2024ApJ...970...93K}, which sets it apart from scintillator-based observations.
In the remainder of this paper, we focus on observations at SK for this reason.

Our group has been dedicated to developing analysis methods specifically targeting neutrinos emitted from the late phase of CCSNe. 
In \citetalias{Suwa_2019}~\citep{Suwa_2019} an analytical approach using differences in neutrino light curves with respect to the initial entropy and the baryon mass of the PNS was developed. 
We systematically calculated the number of neutrinos that can be observed at SK over periods lasting more than 20 seconds, employing the database of~\cite{Nakazato_2013}. 
Those findings indicate that neutrinos can be observed for 30 seconds and longer even when low-mass neutron stars (gravitational mass of $1.20M_\odot$) are generated. 
For high-mass neutron stars (gravitational mass of $2.05M_\odot$), neutrinos can be observed for more than 100 seconds. 
Moreover, we demonstrated that the distribution of neutrino events as a function of backward time, the time measured starting from the last observed event, provides information about the neutron star's mass. 
Hereafter, this method is called the backward-time analysis. 
In \citetalias{Nakazato_2022} \citep{Nakazato_2022}, we investigated the impact of the nuclear equation of state (EOS) on the observed neutrino light curves during the late phase. 
Using the backward-time analysis, the event distribution is distinctly characterized by the underlying EOS.
Indeed, the high-density part of the EOS determines the radius of the resulting PNS and the duration of neutrino observation is longer for PNSs with smaller radii. 
The duration is also affected by the low-density part of the EOS, becoming longer due to the presence of heavy nuclei in the low-density region. 
Furthermore, the average energy of neutrinos becomes higher for an EOS that has heavy nuclei with larger mass numbers due to their large neutrino coherent scattering cross section making them act as thermal insulation near the surface of a PNS. 
Consequently, a new analysis method to extract the time variability of the neutrino average energy was proposed in \citetalias{Nakazato_2022}.

Although these studies have successfully demonstrated the power of their analysis methods to extract physical properties of the PNS, it should be noted that they lack a realistic estimation of experimental backgrounds expected during an actual observation.
Identifying the last observed event is a crucial step for probing EOS-dependent features in the late phase signal, yet this identification requires careful treatment of the backgrounds which become more important as the neutrino rate drops at late times.
Therefore, the present work focuses on a precise treatment of signal neutrino interactions amid background events at SK, evaluating a realistic background contamination at late times. 
We have additionally improved the event generation simulation with regard to previous publications in order to maximize the signal channels available for supernova neutrino detection.
Accordingly, we have developed an analytic method to determine the last observed event for this more realistic observation scenario at SK. 

The structure of this paper is as follows.
Supernova models used in the analysis 
are described in Section~\ref{cap:model}.
Section~\ref{cap:method} describes the simulation of both signal and background and additionally explains the analysis method used to identify the last observed event.
We then simulate the time distribution of the last observed event and demonstrate analysis methods using time information from the cumulative observation in Section~\ref{cap:result}.
Furthermore, the performance of model identification using time information is also discussed. 
Finally, we summarize our findings and conclusions in Section~\ref{cap:conclusion}.

\section{Supernova models and equations of state}\label{cap:model}
\subsection{Numerical Models}\label{cap:pns}
We employ numerical models derived from PNS cooling simulations, as detailed in \citetalias{Suwa_2019} and \citetalias{Nakazato_2022}. 
These simulations, beginning several hundred milliseconds after the core bounce, utilize the general relativistic quasi-static evolutionary code with neutrino diffusion \citep{1994pan..conf..763S}. 
In this framework, neutrino transfer is solved using a multi-group flux-limited diffusion scheme, assuming spherical symmetry. While recent studies have explored the dynamics and neutrino emission of a PNS by solving the full Boltzmann neutrino transport equation under the assumption of axisymmetry—which induces convection \citep{2023ApJ...944...60A}—our approach allows us to model neutrino emission over extended timescales, exceeding tens of seconds, and to examine a broad range of PNS models with varying masses and EOSs.

We utilize PNS models with baryon masses ranging from $M_b=1.40M_\odot$ to $1.86M_\odot$. The initial conditions for these models are provided from supernova cores generated by core-collapse simulations when the shock wave reaches the corresponding mass coordinate. For these simulations, we use the general relativistic neutrino radiation hydrodynamics code \citep{2005ApJ...629..922S}, and adopt progenitors with zero-age main-sequence (ZAMS) masses of $M_{\rm ZAMS}=15M_\odot$ and $40M_\odot$ \citep{1995ApJS..101..181W}. We define the start time of the PNS cooling simulation as $t_{\rm init}$, measured from the core bounce, and focus on the neutrino light curve subsequent to $t_{\rm init}$ in this study. It is important to note that $t_{\rm init}$ varies among the models (refer to Table 1 in \citetalias{Nakazato_2022}). The configurations of the PNS cooling models are detailed extensively in \citetalias{Nakazato_2022}.

In this study, we consider four models for the equation of state (EOS): Shen EOS \citep[][]{Shen_1998,Shen_2011}, LS220 EOS \citep[][]{LS_1991}, Togashi EOS \citep[][]{2017NuPhA.961...78T}, and Furusawa-Togashi EOS \citep[][]{2017JPhG...44i4001F}. While the first three EOS models were utilized in \citetalias{Nakazato_2022}, the Furusawa-Togashi EOS is introduced for the first time in this study and is briefly described later. As in \citetalias{Nakazato_2022}, the PNS models in this study are denoted as xxxYzz, where xxx and zz indicate $M_b$ and $M_{\rm ZAMS}$, respectively. The variable ${\rm Y}$ designates the EOS model: ${\rm S}$, L, T, and F denote the Shen EOS, LS220 EOS, Togashi EOS, and Furusawa-Togashi EOS, respectively. For example, the model 147S15 represents a configuration with $M_b=1.47M_\odot$, $M_{\rm ZAMS}=15M_\odot$, and the Shen EOS. For each EOS, we examine models with $M_b=1.62M_\odot$, which include two configurations corresponding to $M_{\rm ZAMS}=15M_\odot$ and $40M_\odot$, differing in their initial conditions. Importantly, numerical data on neutrino emission for models utilizing the Shen EOS, LS220 EOS, and Togashi EOS are publicly available online~\citep{zenodo_nakazato}.
\subsection{New Models with the Furusawa-Togashi EOS}\label{cap:fteos}
The Furusawa-Togashi EOS is an extended version of the Togashi EOS. The phase diagram of nuclear matter is broadly divided into a uniform phase and a non-uniform phase. While uniform nuclear matter consists mainly of free nucleons and electrons, with some light nuclei potentially included, non-uniform nuclear matter is characterized by the formation of heavy nuclei at lower densities and temperatures. 
The Furusawa-Togashi EOS directly utilizes the uniform matter by~\cite{2013NuPhA.902...53T}, as in the Togashi EOS.
However, they differ in their descriptions of non-uniform matter. In the Togashi EOS, the Thomas-Fermi approximation is adopted to describe the structure of non-uniform nuclear matter, assuming a single representative species of nuclei. In contrast, the Furusawa-Togashi EOS models non-uniform matter as a mixture of multiple nuclear species, assuming nuclear statistical equilibrium.

The difference between the Togashi EOS and the Furusawa-Togashi EOS and its impacts on the PNS cooling have already been investigated in \citet{2023PTEP.2023a3E02S}. 
That result indicates the heavy nuclei of the Furusawa-Togashi EOS smaller mass numbers and represent a smaller fraction of the matter than those of the Togashi EOS in the surface region of the PNS. 
Since heavy nuclei have a large cross section for coherent scattering with neutrinos, which is proportional to the square of the mass number, the neutrino opacity is larger and the outer layer of PNS gets hotter for an EOS with larger mass number nuclei \citep[][]{Nakazato_2018}. 
Therefore,  PNS cooling with the Furusawa-Togashi EOS leads to neutrino emission with a lower average energy and a shorter duration compared to the Togashi EOS.

\begin{figure*}[htbp]
    \centering
       \gridline{
        \fig{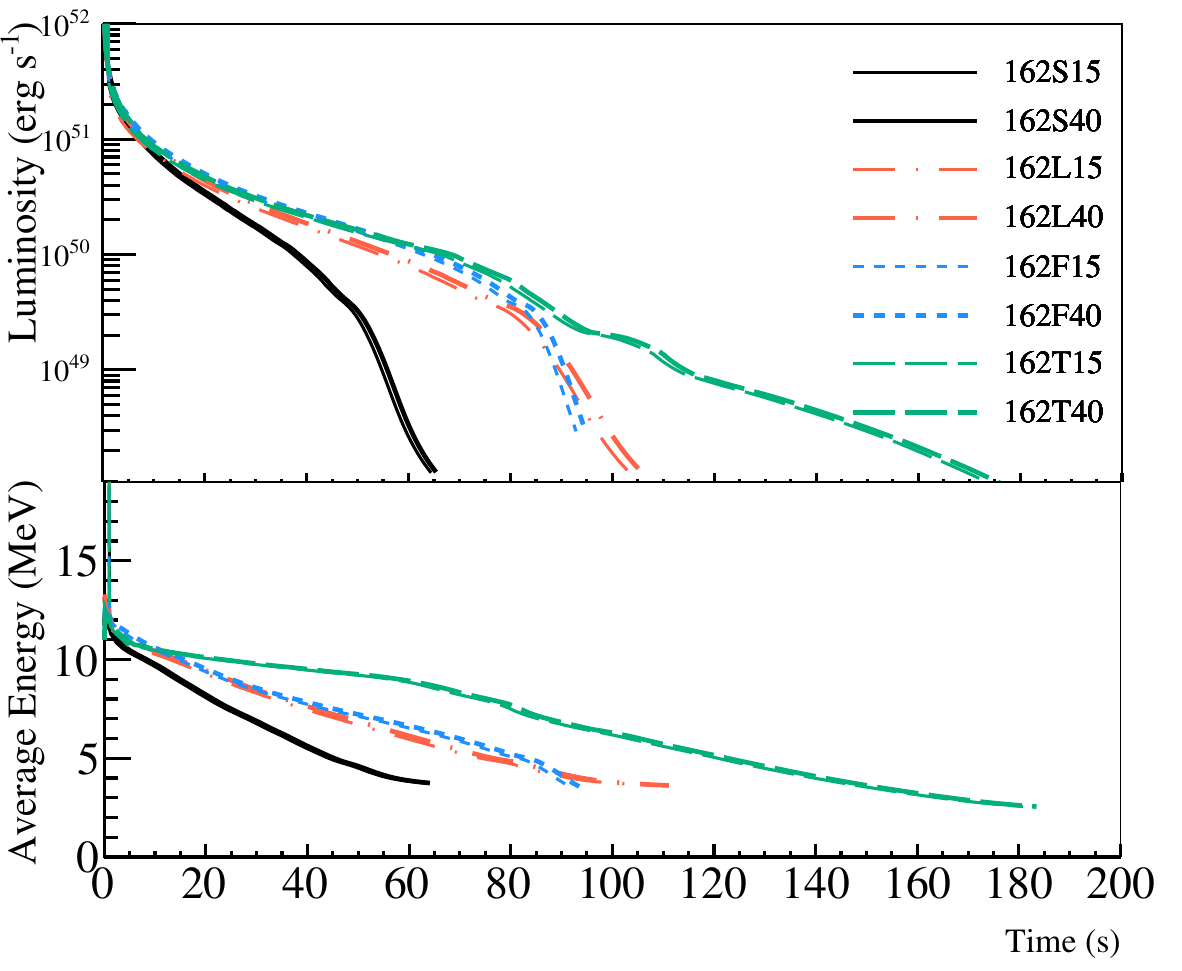}{0.45\textwidth}{}
        \fig{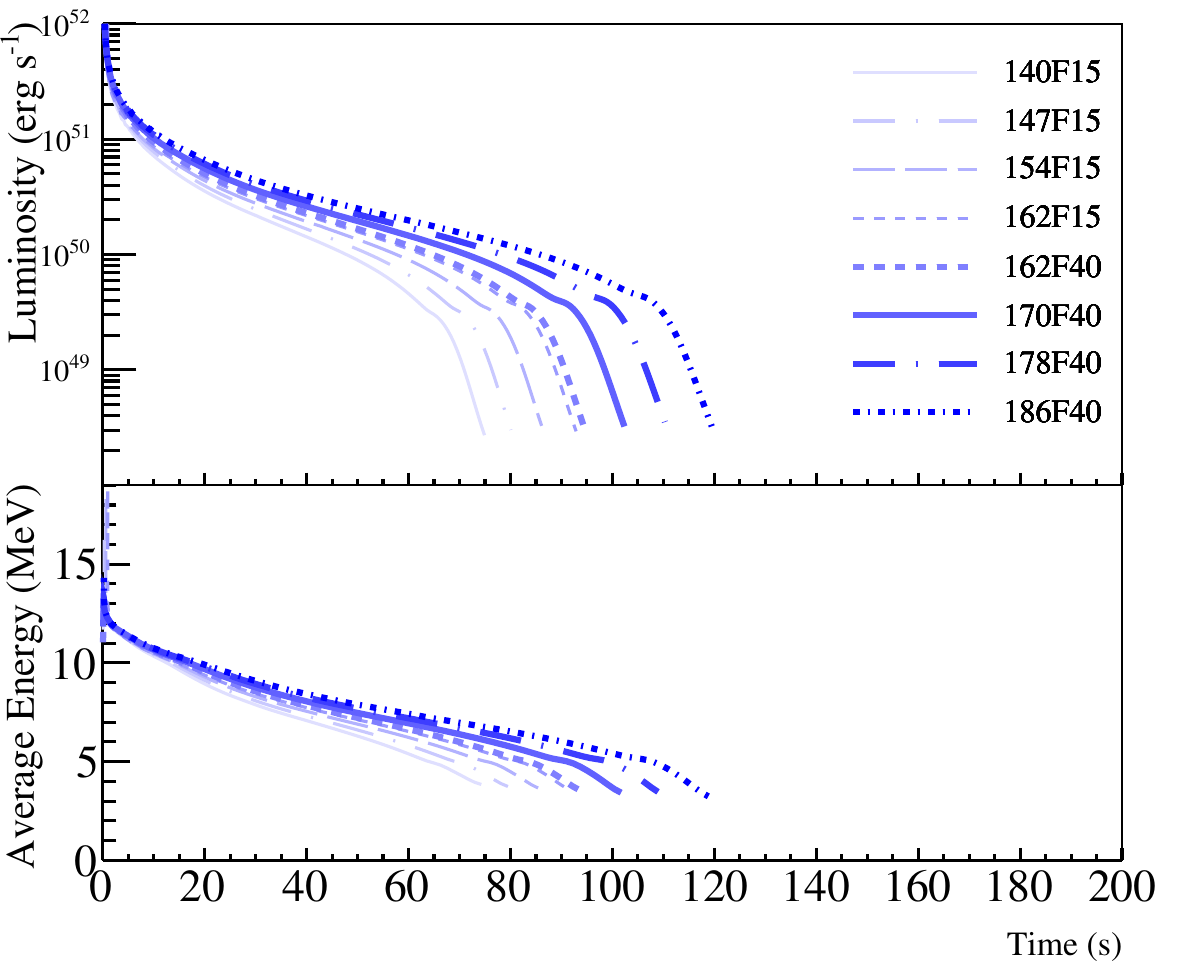}{0.45\textwidth}{}
    }
    \caption{Luminosity (upper) and average energy (lower) of $\bar\nu_e$ emitted during PNS cooling as a function of time after the bounce. The left panel is for PNS models with a baryon mass of $M_b = 1.62M_\odot$, where thin and thick lines correspond to models with $M_{\rm ZAMS} = 15M_\odot$ and $40M_\odot$, respectively, and solid (black), dashed (green), dotted-dashed (red), and dotted (blue) lines correspond to models with the Shen EOS, the Togashi EOS, the LS220 EOS, and the Furusawa-Togashi EOS, respectively. The right panel is for models with the Furusawa-Togashi EOS, where the lines correspond, from bottom to top, to $(M_b, M_{\rm ZAMS}) = (1.40M_\odot, 15M_\odot)$, $(1.47M_\odot, 15M_\odot)$, $(1.54M_\odot, 15M_\odot)$, $(1.62M_\odot, 15M_\odot)$, $(1.62M_\odot, 40M_\odot)$, $(1.70M_\odot, 40M_\odot)$, $(1.78M_\odot, 40M_\odot)$, $(1.86M_\odot, 40M_\odot)$.}
    \label{fig:enelumi}
\end{figure*}

In this paper, PNS cooling with the Furusawa-Togashi EOS is computed for the same PNS mass models in \citetalias{Nakazato_2022}. 
For this purpose, the initial conditions are set identical to those of the Togashi EOS models with the same $M_b$ and $M_{\rm ZAMS}$. 
This is justified by the fact that initially the PNS is composed of uniform nuclear matter, for which the Furusawa-Togashi EOS is identical to the Togashi EOS~\citep[see, also][]{2023PTEP.2023a3E02S}. 
We show the $\bar\nu_e$ luminosity and average energy of the PNS cooling models in Figure~\ref{fig:enelumi}. The left panel focuses on the dependence on the EOS and $M_{\rm ZAMS}$ for models with $M_b = 1.62M_\odot$. Comparing models with $M_{\rm ZAMS} = 15M_\odot$ and $40M_\odot$, we find that the difference in the initial condition has only a minor impact on the neutrino signal, as reported in \citetalias{Suwa_2019}. As for the EOS dependence, PNS models with the Shen EOS have a shorter timescale than those with other EOSs. This is because the Shen EOS results in a larger neutron star radius and less compact PNS configurations lead to more rapid cooling. Since the Togashi EOS and the Furusawa-Togashi EOS share the high-density region, which primarily determines the neutron star radius, their PNS models exhibit a similar neutrino light curve during the Kelvin-Helmholtz cooling phase \citep[][]{2019ApJ...878...25N,2020ApJ...891..156N}, which corresponds to the period until the $\bar\nu_e$ luminosity drops to approximately $10^{50}~{\rm erg}~{\rm s}^{-1}$. As shown in the right panel of Figure~\ref{fig:enelumi}, models with a higher PNS mass have longer neutrino emission.
The luminosity of the Togashi EOS declines slowly because this model has a high abundance of heavy nuclei within the PNS.


\section{Framework for Mock Data Construction}
\begin{figure*}[tbp]
\centering
\tikzstyle{line} = [draw, -stealth]
\tikzstyle{block} = [rectangle, draw, rounded corners, minimum width=3cm, minimum height=1cm, align=center]
\tikzstyle{cloud} = [ellipse, draw, minimum height=1cm, minimum width=2cm, align=center]
\begin{tikzpicture}[node distance = 1.8cm, auto]
    \definecolor{customrgb}{HTML}{AADDDD}
    \definecolor{customgrn}{HTML}{E9FC9D}
    \node [block, fill=customrgb] (sksnsim) {\hyperlink{eventgen}{
    Supernova signal event generation
    }};
    \node [block, fill=customrgb, below of=sksnsim] (eventselection) {\hyperlink{skbackground}{Event selection}};
    \node [block, fill=customgrn, below of=eventselection, yshift=-0.5cm, xshift=0cm] (events) {Supernova signal event};
    \node [block, fill=customrgb, below of=events, yshift=-0.5cm, xshift=1.97cm] (tlast) {\hyperlink{tlast}{$T_{\text{last}}$ determination}};
    \node [block, fill=customgrn, right of=events, xshift=2.2cm] 
    (background) {SK background event};
    \node [cloud, left of=sksnsim, xshift=-3.4cm] (direction) {Supernova direction};
    \node [cloud, right of=sksnsim, xshift=3.2cm] (fullvolume) {$D_{\rm{wall}}>0~\rm{cm}$ (ID)};
    \node [cloud, above of=sksnsim, yshift=-0.5cm, xshift=-3cm] (10kpc) {10 kpc};
    \node [cloud, right of=eventselection, xshift=2.2cm, text width=3cm, align=left] (eventrate) {
    \begin{tabular}{l}
    $E > E_{\rm{th}}$ \\
    $D_{\text{wall}} > 200 \text{ cm (FV)}$ \\
    Spallation cut
    \end{tabular}
    };
    \node [cloud, left of=eventselection, xshift=-3.5cm, text width=4.3cm, align=left] (evselcriteria) {
        \begin{tabular}{l}
            $E > E_{\rm{th}}$ \\
            $D_{\text{wall}} > 200 \text{ cm (FV)}$ \\
            Randomly reduces events by 20\%
        \end{tabular}
    };
    \path [line] (sksnsim) -- (eventselection);
    \node at ($(events)!0.5!(background)$) {+};  
    \path [line] (direction) -- (sksnsim);
    \path [line] (fullvolume) -- (sksnsim);
    \path [line] (10kpc) -- (sksnsim);
    \path [line] (eventrate) -- (background);
    \path [line] (evselcriteria) -- (eventselection);
    \node [draw, rounded corners, fit=(events)(background), inner sep=0.5cm] (mcsample) {};
    \node[anchor=north] at (mcsample.north) {mock sample};
    \path [line] (eventselection) -- (events);
    \path [line] (mcsample) -- (tlast);
\end{tikzpicture}

\caption{Schematic diagram of the $T_{\rm{last}}$ determination method. 
$D_{\rm{wall}}$ and $E_{\rm{th}}$ represent the distance from the wall of the inner tank, and the energy threshold, respectively. A 20\% random reduction is applied to simulate the expected signal loss due to spallation background rejection. 
The blue blocks indicate the flow of the $T_{\rm{last}}$ determination method, while the green blocks represent the generated events.}
\label{fig:sim_flow}
\end{figure*}
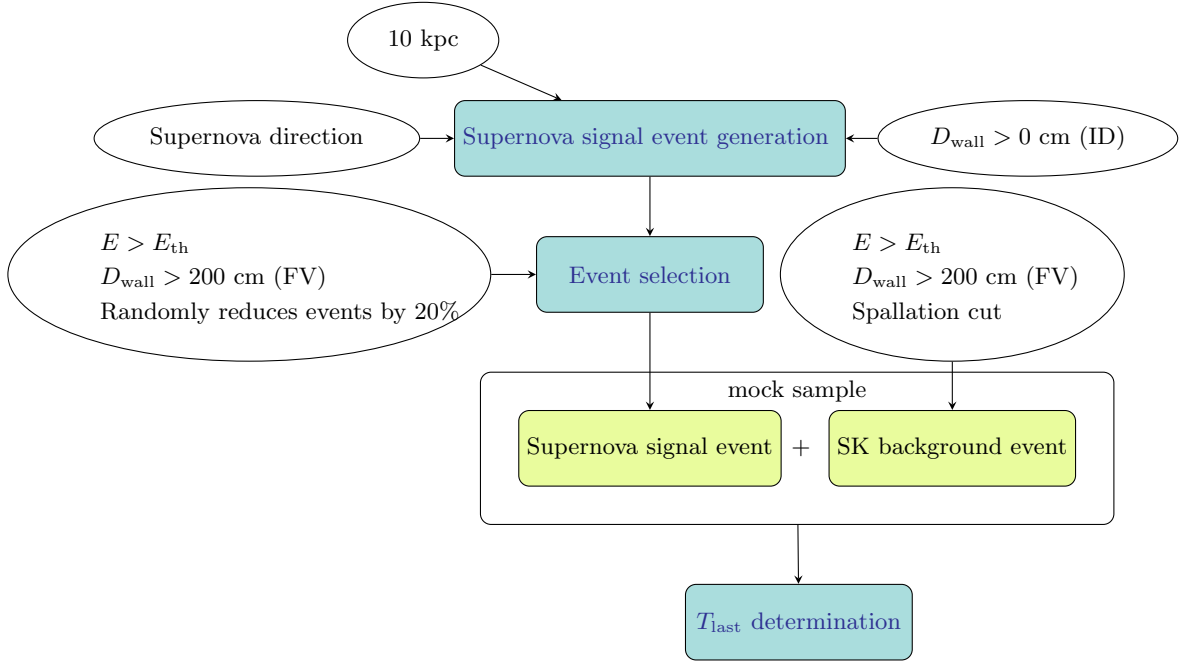

In order to develop the analysis method for supernova neutrinos, we generate mock samples for both signal and background events.
In Figure \ref{fig:sim_flow}, we present a schematic diagram summarizing the simulation flow. 
First, we perform Monte Carlo (MC) simulations to generate supernova signal events and separately generate background events.
Next, for each such realization, we combine the supernova signal and background events to construct mock observation samples.
Finally, as part of the simulation process, the method described in Section~\ref{cap:technique} is applied to select the last observed event.
This entire procedure is repeated 1000 times to evaluate statistical fluctuations in the result.
The results of this procedure are presented in Section~\ref{cap:result}.

\label{cap:method}
\subsection{Supernova Signal}
\label{cap:sk}
\hypertarget{eventgen}{}
We simulate the supernova neutrino signal using SKSNSim (Super-Kamiokande SuperNova Simulator;~\cite{zenodo_nakanishi})
, a tool capable of simulating all types of neutrino interactions in water Cherenkov detectors such as SK~\citep{2024ApJ...965...91N}.\footnote{
Our group has reported similar studies with simplified detector responses~\citep{2022ApJ...934...15S, 2023ApJ...954...52H, 2025ApJ...980..117S}, demonstrating the capability for PNS parameter estimation. The code for these studies is publicly available at~\citep{zenodo_harada}.}

The neutrino interactions observed in SK for neutrino energies below $100~\rm{MeV}$ are categorized into the following four types: 
\begin{subequations}
    \begin{align}
        &\text{Inverse Beta Decay (IBD):} \nonumber \\ 
        &\quad \nuebar + p \to n + e^+, \\ 
        &\text{Electron Scattering (ES):} \nonumber \\ 
        &\quad \nue/\nuebar/\nu_x/\nuxbar + e^- \to \nue/\nuebar/\nu_x/\nuxbar + e^-, \\ 
        &\text{Charged-Current reaction with oxygen ($\mathrm{^{16}O}$ CC):} \nonumber \\ 
        &\quad \nue/\nuebar + \mathrm{^{16}O} \to e^-/e^+ + \mathrm{^{16}F/^{16}N}, \\ 
        &\text{Neutral-Current reaction with oxygen ($\mathrm{^{16}O}$ NC):} \nonumber \\ 
        &\quad \nue/\nuebar/\nu_x/\nuxbar + \mathrm{^{16}O} \to p/n + \gamma + \mathrm{^{15}N/^{15}O}.
    \end{align}
\end{subequations}
The dominant channel in the relevant energy region at SK is the IBD reaction,  whose event rate is written as: 
\begin{equation}
    \frac{dN(E_\nu,t)}{dt}=\int_{0~\rm{MeV}}^{100~\rm{MeV}}dE_\nu N_p\sigma(E_\nu)\frac{d\phi(E_\nu,t)}{dE_\nu}
    \label{eq:eve}
\end{equation}
where $N_p$ is the number of free protons. 
The energy range is set from 0 to 100 MeV in SKSNSim, which is associated with the energy range of supernova neutrinos. 
$d\phi(E_\nu,t) / dE_\nu$ is the neutrino number spectrum of $\Bar{\nu}_e$ in units of $[\rm{s^{-1}~cm^{-2}}]$, which is estimated as: 
\begin{equation}
    \frac{d\phi(E_\nu,t)}{dE_\nu} = \frac{1}{4\pi D^2}\frac{d^2N_\nu(E_\nu,t)}{dE_\nu dt},
    \label{eq:spe}
\end{equation}
where $D$ is the distance between the detector and the supernova.  
A distance of $D = 10~\rm{kpc}$ is assumed in this paper.
The quantity $d^2N_\nu(E_\nu,t)/dE_\nu dt$ represents the luminosity spectrum of neutrinos emitted by a single supernova, for which we use the models described in Section~\ref{cap:model}. 
The cross section for IBD is denoted as $\sigma(E_\nu)$, following the formulation provided in \cite{Strumia_2003}. 
Neutrino event rate spectra for other interactions are calculated in the same manner (Equation~\ref{eq:eve}) but with the corresponding target and neutrino flavors.
For more details, we refer the reader to~\cite{2024ApJ...965...91N}.
\subsection{Background Sources and Reduction Strategies}
\hypertarget{skbackground}{}
In this study, to incorporate realistic observational conditions, we incorporated the in-situ background estimation from \cite{Mori_2022}.
This step is an essential component for developing a realistic method to determine the last observed event.

In the search for supernova neutrinos, there are several types of background, such as radioimpurities, spallation background, decay electrons from muons, and atmospheric neutrinos.
In particular, the reduction of background from radioimpurities and spallation products is especially important in the analysis of neutrinos emitted from the late phase of CCSNe.

In the case of radioimpurities, such backgrounds mainly originate from the detector materials and the surrounding rock of the SK tank.
In order to reduce radioactive backgrounds a fiducial volume (FV) cut is applied which excludes events occurring within $2~\rm{m}$ of the tank wall.
As a result, the tank volume is reduced from the entire inner detector (ID) volume of $32.5~\rm{kton}$ to $22.5~\rm{kton}$.
Analysis in SK typically focuses on events within this FV.

The ``spallation background'' arises from the beta-decays of the radioisotopes generated by the spallation of oxygen nuclei in the tank by cosmic-ray muons.
Such spallation creates hadronic showers that result in secondary hadrons undergoing hadronic interactions with detector nuclei to produce various radioisotopes.
The observable particles appearing in the final, such as beta and gamma rays, have energies ranging from a few MeV to $\sim$20 MeV, which overlaps with the energy range for supernova neutrinos. 
 Consequently, these events can contribute to the background during supernova neutrino observation.

\begin{figure}[htbp]
    \centering
    \includegraphics[scale=0.45]{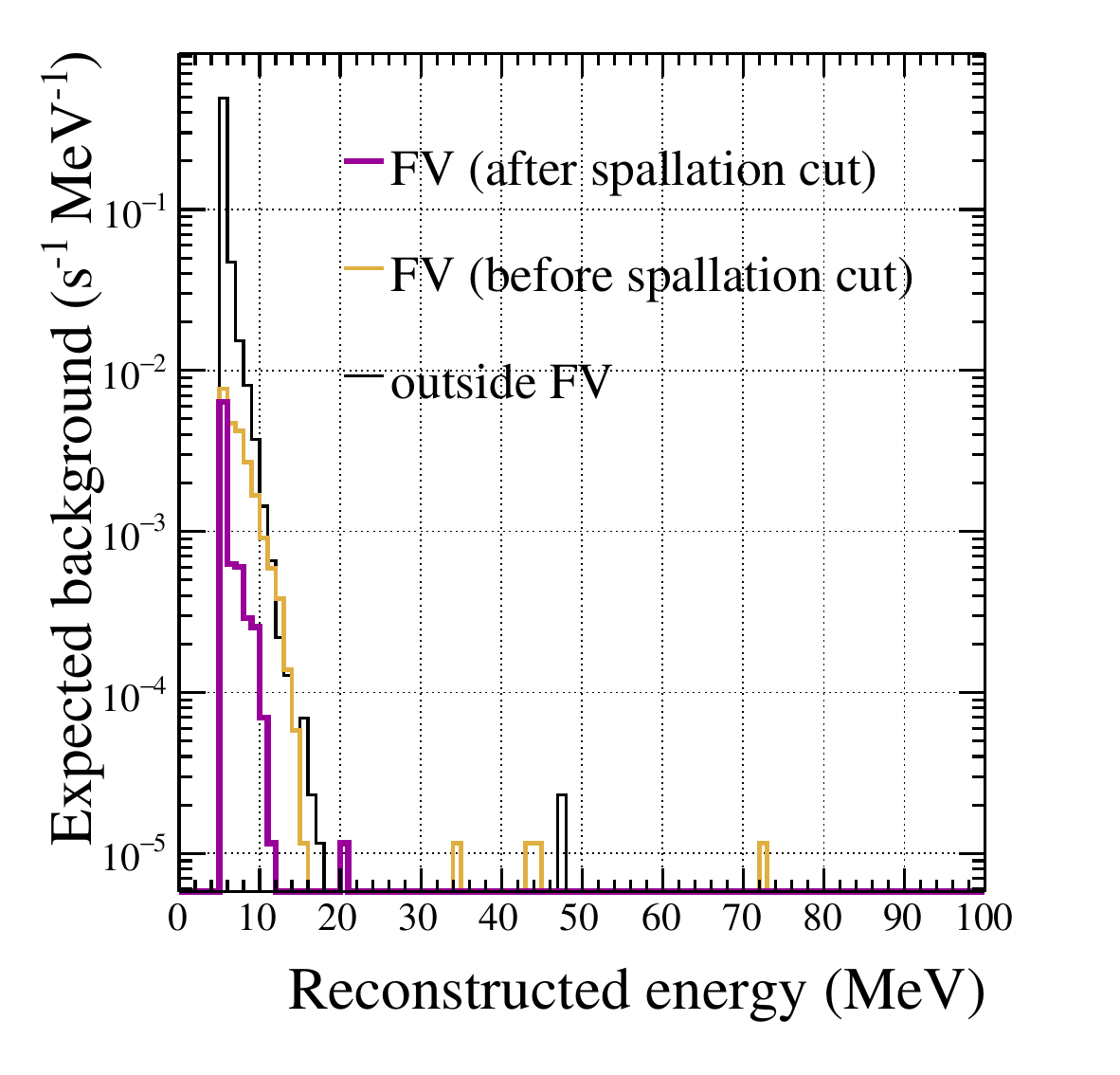}
    \caption{Background rate for the FV (orange), FV with spallation cut (violet), and outside FV (black) samples in SK.  Background rates are generated according to~\cite{Mori_2022}. In the present analysis, we use the background rate above $5~\rm{MeV}$ following that study.}
    \label{fig:bkgID_FV}
\end{figure}

Spallation events are spatially and temporally correlated with the trajectory of muon passage.
Super-Kamiokande therefore introduces a  ``spallation cut'' that utilizes such correlations to remove approximately 90\% of all spallation events~\citep{2024PhRvD.110c2003L}.
However, this spallation cut also removes 20\% of supernova neutrino signal events~\citep{2024PhRvD.109i2001A}.
Figure~\ref{fig:bkgID_FV} shows the expected background rates in SK for three cases: within the FV, within FV after applying a spallation cut, and outside FV. 
It shows that the FV cut removes $96.1\%$ of background events in the ID and that $64.7\%$ of the remaining background is eliminated by the spallation cut.
Consequently, the final background rate for events with energies above $5~\rm{MeV}$ is $ 8.2\times10^ {-3}~(\rm{s^{-1})}$.

\begin{figure*}[htb!]
    \gridline{
        \fig{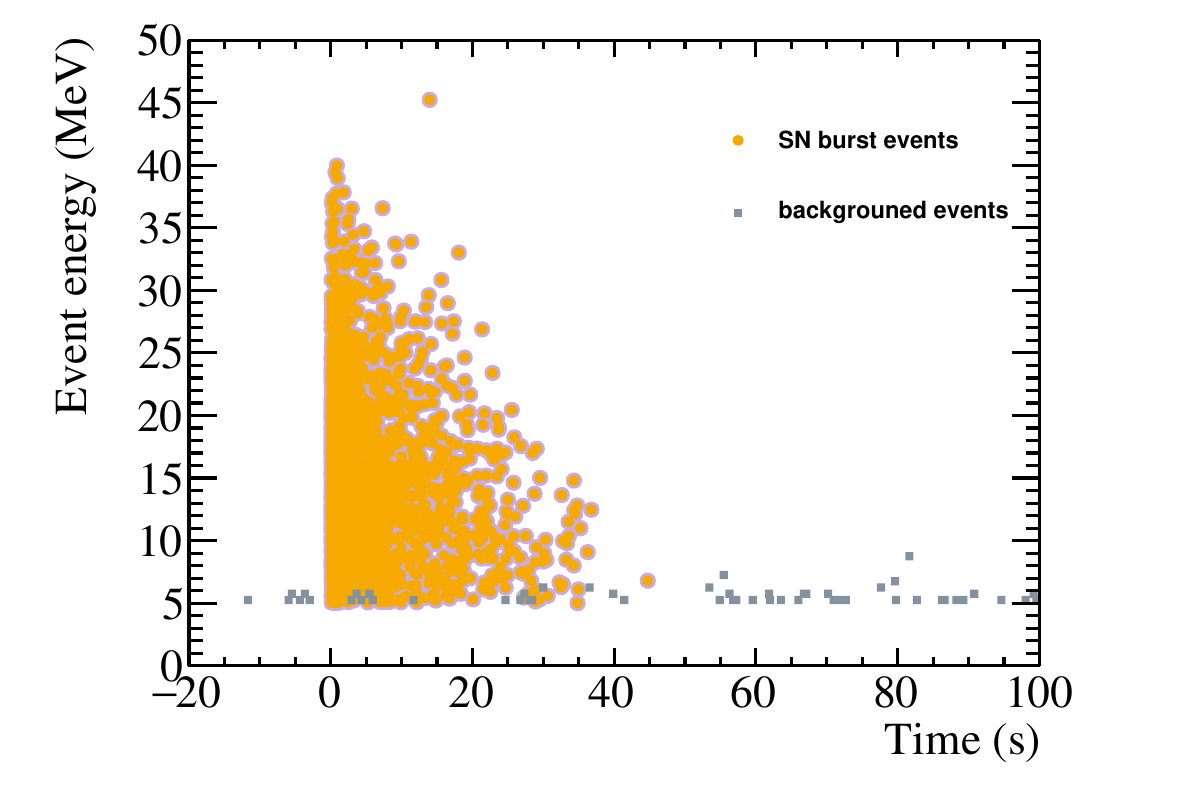}{0.53\textwidth}{(a) Entire inner detector (ID)}
        \fig{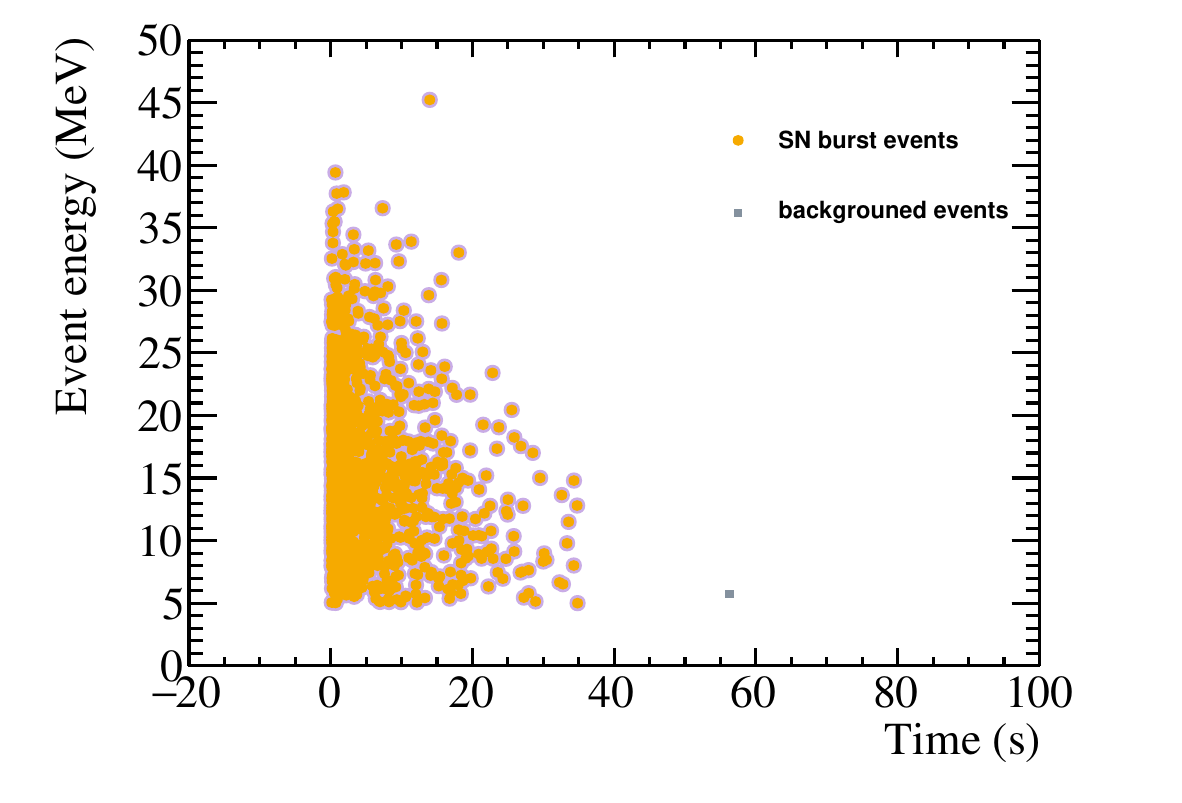}{0.53\textwidth}{(b) After reduction}
    }
    \caption{An example scatter plot showing the visible energy of the observed events in SK as a function of time from the supernova explosion based on a single simulation. Signal events, represented by the orange points, and background events, represented by the gray points, are shown for the case with an energy threshold of 5 MeV. For signal events we use the Shen EOS with a $1.40M_{\odot}$ PNS mass. Panel (a) is assumed to be the full volume observation, while panel (b) applies a spallation cut within the fiducial volume.
    }
    \label{fig:scat}
\end{figure*}

In Figure \ref{fig:scat}, we present an example scatter plot showing the true neutrino energy of MC as a function of time for a single MC simulation. 
These signal events correspond to the model featuring the Shen EOS and a $1.40M_{\odot}$ PNS mass. 
Figure~\ref{fig:scat} (a) and (b) respectively represent the event distribution in the entire ID case and that with both the FV and spallation cuts applied.
In the present analysis, both the FV cut and the spallation cut are applied to ensure a low-background environment.

Averaged over all models, approximately 69\% of the signal events remain after applying the FV cut, and about 55\% remain after the additional spallation cut.
The specific interaction channel does not affect the signal efficiency.
\subsection{Technique to Determine the Last Observed Event}
\label{cap:technique}
\begin{figure}[ht]
   \begin{center}
      \includegraphics[scale = 0.4]{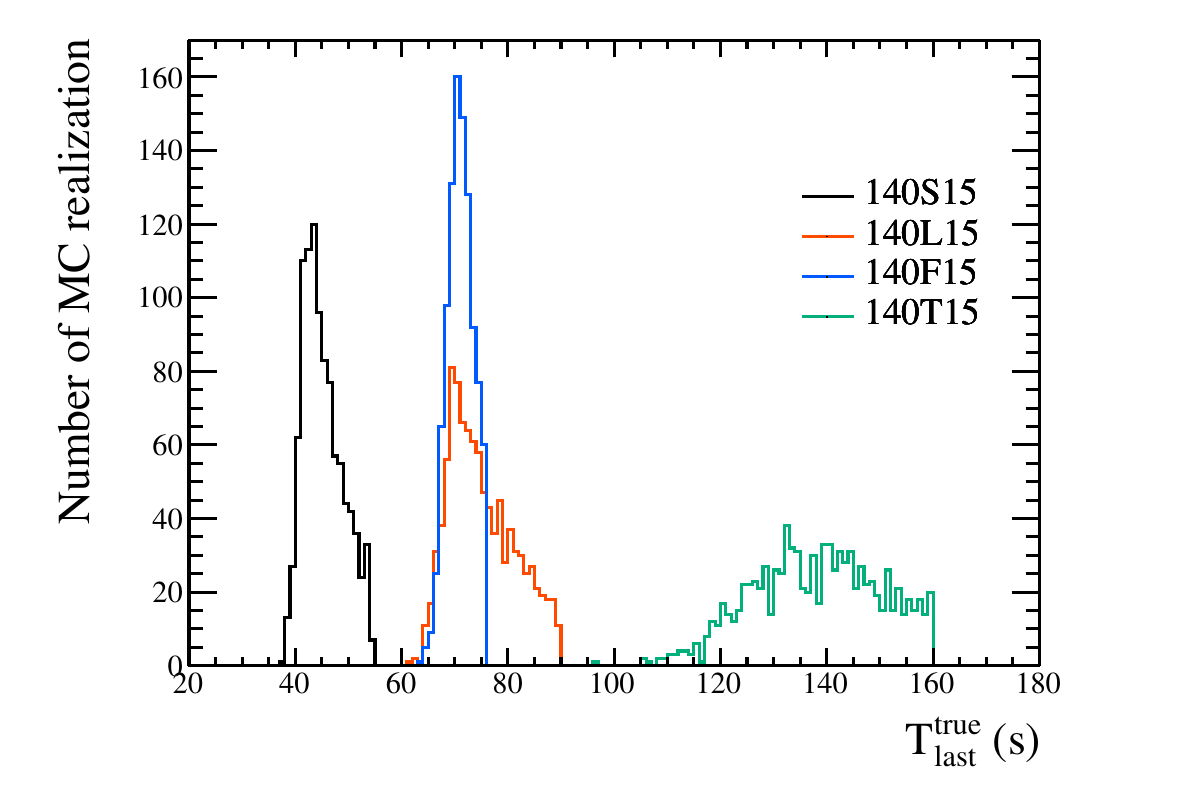}
      \caption{$T^{\rm{true}}_{\rm{last}}$ distribution for PNS models with a baryon mass of $M_b = 1.40M_{\odot}$, represented as follows: black for the Shen EOS, red for the LS220 EOS, blue for the Furusawa-Togashi EOS, and green for the Togashi EOS case. This plot is made from 1000 MC realizations; the vertical axis shows the number of MC realizations per second. We assume observations from supernova at a distance of 10~kpc.}
   \label{fig:Tlast_true_140}
   \end{center}
\end{figure}

\hypertarget{tlast}{}
In this section, we describe how to determine the last observed event.
To begin with, by performing 1000 SN event generations using SKSNSim, we investigate the distribution of the last observed event under the assumption of no background and no cuts.
The detection time of the last observed event, generated without any cuts or background, is denoted as $T^{\rm{true}}_{\rm{last}}$ and its resulting distribution is shown in Figure~\ref{fig:Tlast_true_140}. 
Some of the $T^{\rm{true}}_{\rm{last}}$ distributions are terminated at the right-edge, because the numerical simulation covers a finite time range.
The Shen EOS exhibits the shortest $T^{\rm{true}}_{\rm{last}}$, while the Togashi EOS features the longest.
Additionally, the Togashi EOS has a broader distribution. 
The $T^{\rm{true}}_{\rm{last}}$ behavior is attributed to the luminosity evolution; $T^{\rm{true}}_{\rm{last}}$ tends to be shorter for a rapid decline in luminosity, whereas a slower luminosity decline results in a longer $T^{\rm{true}}_{\rm{last}}$, as shown in Figure~\ref{fig:enelumi}.

\begin{figure*}[htbp]
\centering
\pgfplotsset{compat=1.17}
\begin{tikzpicture}[scale=0.8]
    \definecolor{customgrn}{HTML}{009999}
    \draw[->,thick] (-1.2,0) -- (11.5,0) node[right, font=\Large] {$T$ (s)};
    \draw[->,thick] (-0.7,-0.5) -- (-0.7,4);
    \draw[dashed, red] (-0.7,0.8) -- (11.5,0.8) node[right] {$E_{\rm th}=8$ MeV};
    \node[rotate=90, font=\Large] at (-1.2,2) {$E$ (MeV)};
    \draw[<->,blue] (0,-0.2) -- (1.5,-0.2);
    \node at (0,-0.7) {$T=0$};
    \draw[<->,blue] (0.4,-0.4) -- (1.9,-0.4) node[right] {$T_{\rm wid}=5$ s};
    \draw[<->,blue] (6.3,-0.2) -- (7.8,-0.2);
    \node at (6.3,-0.5) {$T_{\rm last}$};  
    \foreach \i/\height in {0/3, 0.4/2.5, 0.9/1.2, 1.3/3.8, 2.1/2, 2.6/2.4, 3.3/2.8, 3.7/1.5, 4.2/1, 4.6/1.7, 5.3/2.1, 6.3/1.3} {
        \draw[customgrn, fill=customgrn!50] (\i-0.1,0) rectangle (\i+0.1,\height);
    }
    \foreach \i/\height in {1.7/0.65, 5.7/0.5, 8.3/1.4, 11/0.6} {
        \draw[customgrn] (\i-0.1,0) rectangle (\i+0.1,\height);
    }
    \foreach \i/\height in {4.2/1, 10/1.2} {
        \draw[fill=gray!50] (\i-0.1,0) rectangle (\i+0.1,\height);
    }    
    \draw[customgrn, fill=customgrn!50] (8-0.1,3.2-0.1) rectangle (8+0.1,3.2+0.1);
    \node[customgrn, font=\large, right] at (8.2,3.2) {\textbf{Signal (SN~$\nu$)}};
    \draw[fill=gray!50] (8-0.1,2.6-0.1) rectangle (8+0.1,2.6+0.1);
    \node[font=\large, right] at (8.2,2.6) {\textbf{Background}};
    \draw[red, decorate, decoration={brace, amplitude=10pt, mirror}, yshift=-2pt] (6.6,3.15) -- (-0.3,3.15);
    \node[font=\Large] at (3.15,3.8) {$N$};
\end{tikzpicture}
    \caption{Schematic diagram of $T_{\rm{last}}$ determination method. It represents supernova neutrino and background events plotted over time, with their corresponding energy values represented on the vertical axis. Green box shows supernova neutrino events whose energy is more than $E_{\rm{th}}$ and the black boxes indicate background events. White boxes with green outlines show supernova neutrino events that are excluded because their energy is below a $E_{\rm{th}}$ or their time difference from $T_{\rm{last}}$ exceeds $T_{\rm{wid}}$ .}
    \label{fig:last_image}
\end{figure*}
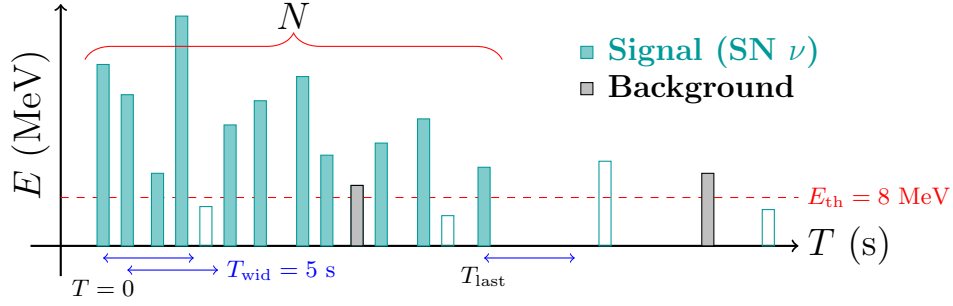

In a realistic analysis, the time of the last observed signal needs to be determined amid possible background contamination.
Therefore, we develop a method for determining the last observed event based on two key parameters: the time width ($T_{\rm{wid}}$), which defines a time window for event selection, and an energy threshold ($E_{\rm{th}}$) which is applied to individual events to reject low-energy background events. 
The detection time of the last observed event selected using the method is denoted as $T_{\rm{last}}$.
Figure \ref{fig:last_image} shows a schematic diagram of how $T_{\rm{last}}$ is determined. 
The time $T$ is measured relative to the first supernova neutrino event observed in SK ($T=0$). 
In order to determine $T_{\rm{last}}$, events with $T \geq 0$ events and with energies a given threshold $E_{\rm{th}}$ are counted within a sliding interval with a given width $T_{\rm{wid}}$.  
The interval slides forward in time until events are no longer detected. 
The time of the latest event in last interval with events is then defined as $T_{\rm{last}}$.


In the following analysis we aim to choose $T_{\rm{wid}}$ and $E_{\rm{th}}$ such that the event defining $T_{\rm{last}}$ is inconsistent with the measured background at more than 5$\sigma$ significance.
To achieve this condition we use the background rate in the FV after applying the spallation cut described above (see Figure~\ref{fig:bkgID_FV}). 
A suitable combination of $T_{\rm{wid}}$ and an $E_{\rm{th}}$ is determined based on Poisson statistics assuming only the measured background rate to establish the required detection significance. 
We note that the background is more easily rejected with smaller values of $T_{\rm{wid}}$ and larger values of $E_{\rm{th}}$. 
In contrast, most signal events are likely to survive even under these conditions due to their concentration within tens of seconds and relatively high energies.
Some combinations of $T_{\rm{wid}}$ and $E_{\rm{th}}$ satisfying the detection requirements above are summarized in Table~\ref{table:t-Eth}.


\begin{table}[hbtb]
    \caption{Pairs of time width and energy threshold that provide no background events at greater than $5\sigma$ for FV events. }
    \label{table:t-Eth}
    \centering
    \begin{tabular}{cc}
         \hline
        Time width ($T_{\rm{wid}}$) [sec] & Energy threshold ($E_{\rm{th}}$) [MeV]\\
         \hline \hline
          1 & 6.0 \\
          5 & 8.0 \\
          10 & 9.5 \\
          20 & 10.0 \\
         \hline
    \end{tabular}
\end{table}
\begin{figure*}[htbp]
    \centering
    \gridline{
        \fig{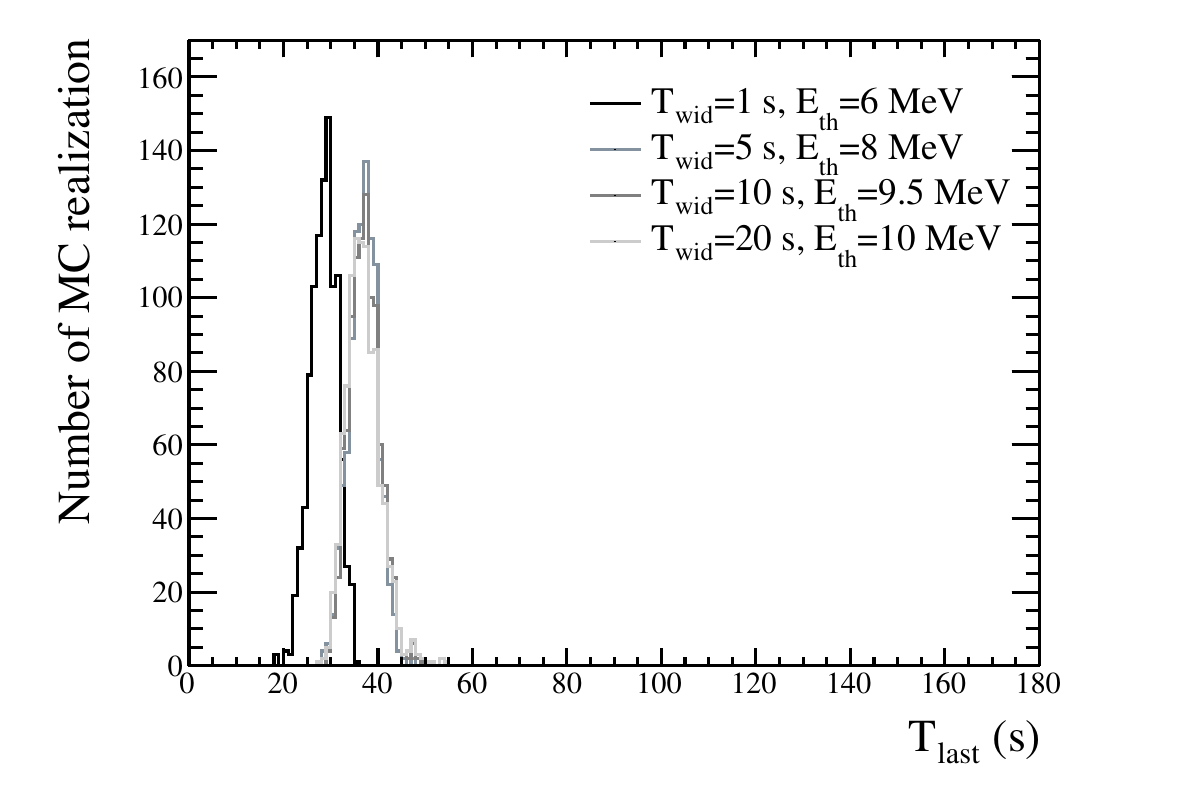}{0.45\textwidth}{(a) Shen EOS }
        \fig{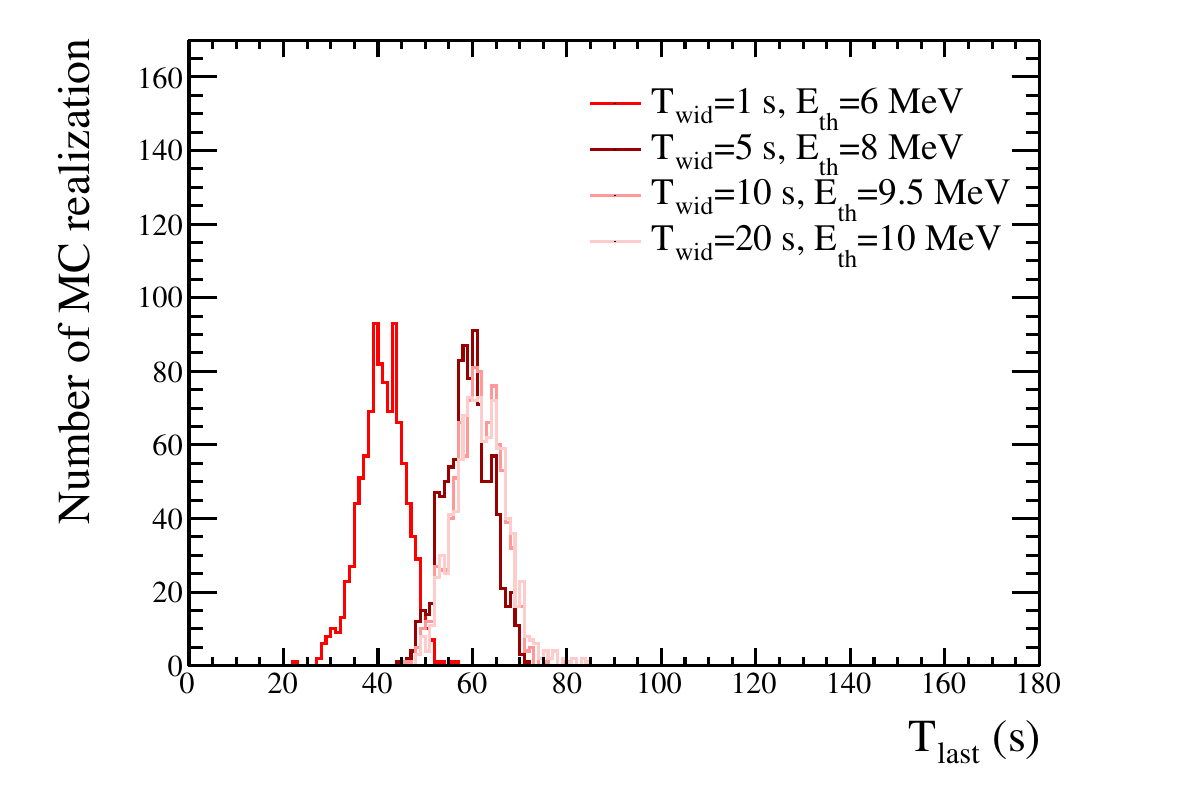}{0.45\textwidth}{(b) LS220 EOS }
    }
    \gridline{
        \fig{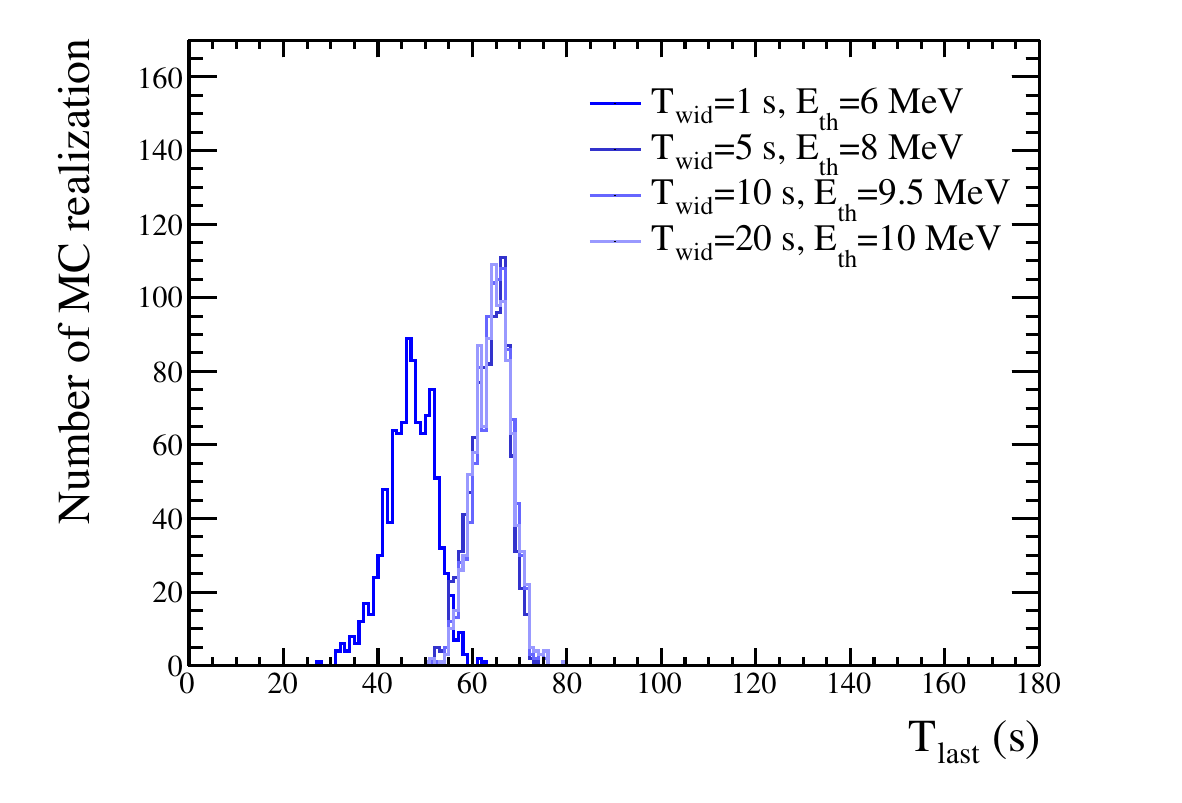}{0.45\textwidth}{(c) Furusawa-Togashi EOS }
        \fig{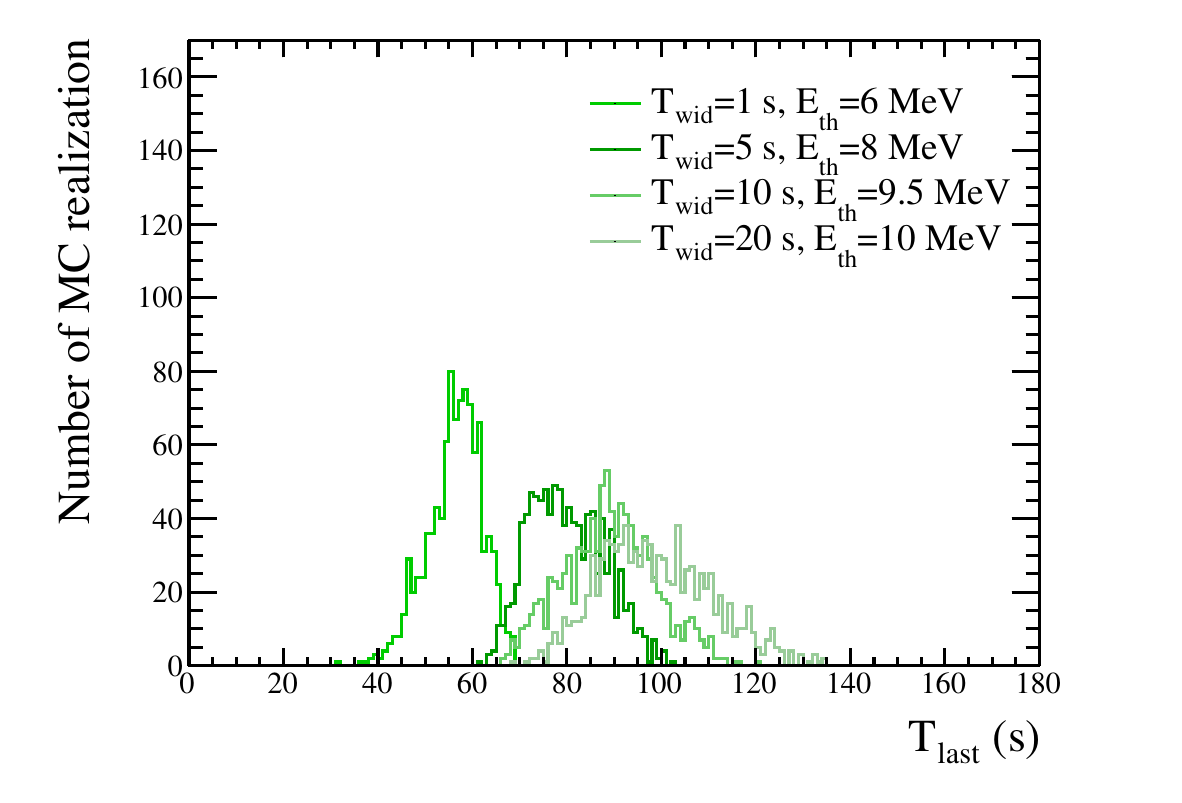}{0.45\textwidth}{(d) Togashi EOS }
    }
    \caption{The $T_{\rm{last}}$ distribution for each combination of $T_{\rm{wid}}$ and $E_{\rm{th}}$. Panel (a), (b), (c), and (d) correspond to Shen EOS, LS220 EOS, Furusawa-Togashi EOS, and Togashi EOS, respectively. Here, the PNS models with mass of $M_b=1.40M_{\odot}$ and $M_{\rm{ZAMS}}=15M_\odot$ are shown.}
    \label{fig:tlast_twid_eth}
\end{figure*}

Figure~\ref{fig:tlast_twid_eth} shows the $T_{\rm{last}}$ distributions selected based on $T_{\rm{wid}}$ and $E_{\rm{th}}$ for four models.  
For $T_{\rm{wid}} \ge 5~\rm{sec}$ the distributions are largely consistent except for the Togashi EOS where more variation is seen.
Note that reducing $E_{\rm{th}}$ in general increases the total number of detected events, improving the statistical reliability. 
In particular, since the average neutrino energy decreases in the late emission phase, lowering $E_{\rm{th}}$ allows for the detection of more events from that phase.
Therefore, we select $T_{\rm{wid}}=5~\rm{sec}$ and $E_{\rm{th}}=8.0~\rm{MeV}$ for the analysis presented below.
As a result of this energy cut, the signal efficiency remains approximately $47\%$ for all supernova models, while about 98\% of background events are rejected.
\section{Analysis Results and Their Implications}
\label{cap:result}
In this section, we show the $T_{\rm{last}}$ distribution determined from mock samples and cumulative event distribution starting from $T_{\rm{last}}$ as time origin. 
The results presented here are all produced under the proper background consideration from the previous section.
Model discrimination using the $T_{\rm{last}}$ information is also discussed.
\begin{table}[hbtb]
    \caption{The average of $T_{\rm{last}}$, and the time difference between the $500\text{-th}$-to-last and $1000\text{-th}$-to-last event ($T_{1000}-T_{500}$) for each model. The error shows the $1\sigma$ range of the distribution.}
    \centering
    \begin{tabular}{c|cc}
        \hline
         Model & $T_{\rm{last}}$ (sec.) & $T_{1000}-T_{500}$ (sec.)\\
         & &\\
        \hline
        \hline
        140S15 & $37.0 \pm 3.0$ & $3.16 \pm 0.22$\\
        147S15 & $39.9 \pm 3.2$ & $3.78 \pm 0.24$\\
        154S15 & $43.4 \pm 3.3$ & $4.55 \pm 0.27$\\
        162S15 & $47.1 \pm 3.6$ & $5.34 \pm 0.29$\\
        162S40 & $48.1 \pm 3.6$ & $5.43 \pm 0.29$\\
        170S40 & $52.4 \pm 3.9$ & $6.19 \pm 0.32$\\
        178S40 & $57.1 \pm 3.9$ & $6.91 \pm 0.35$\\
        186S40 & $61.8 \pm 4.4$ & $7.67 \pm 0.39$\\
        \hline
        140L15 & $59.0 \pm 4.8$ & $5.86 \pm 0.35$\\
        147L15 & $64.7 \pm 5.0$ & $6.93 \pm 0.40$\\
        154L15 & $70.0 \pm 5.4$ & $8.04 \pm 0.43$\\
        162L15 & $76.2 \pm 5.8$ & $9.17 \pm 0.47$\\
        162L40 & $78.4 \pm 5.9$ & $9.41 \pm 0.48$\\
        170L40 & $86.7 \pm 6.1$ & $10.74 \pm 0.56$\\
        178L40 & $95.2 \pm 6.9$ & $12.24 \pm 0.61$\\
        186L40 & $105.0 \pm 7.5$ & $13.82 \pm 0.72$\\
        \hline
        140F15 & $63.9 \pm 4.0$ & $7.06 \pm 0.38$\\
        147F15 & $64.7 \pm 4.0$ & $8.08 \pm 0.44$\\
        154F15 & $75.1 \pm 4.2$ & $9.25 \pm 0.49$\\
        162F15 & $81.0 \pm 4.2$ & $10.49 \pm 0.54$\\
        162F40 & $83.0 \pm 4.4$ & $10.72 \pm 0.56$\\
        170F40 & $90.3 \pm 4.6$ & $12.13 \pm 0.61$\\
        178F40 & $98.5 \pm 4.7$ & $13.64 \pm 0.69$\\
        186F40 & $107.2 \pm 4.8$ & $15.30 \pm 0.76$\\
        \hline
        140T15 & $80.5 \pm 7.7$ & $9.76 \pm 0.57$\\
        147T15 & $86.7 \pm 8.2$ & $11.40 \pm 0.62$\\
        154T15 & $93.1 \pm 8.0$ & $13.25 \pm 0.71$\\
        162T15 & $100.9 \pm 8.4$ & $15.08 \pm 0.79$\\
        162T40 & $102.2 \pm 8.3$ & $15.35 \pm 0.83$\\
        170T40 & $111.3 \pm 8.6$ & $17.47 \pm 0.87$\\
        178T40 & $120.2 \pm 8.8$ & $19.59 \pm 0.94$\\
        186T40 & $129.8 \pm 9.1$ & $21.76 \pm 1.03$\\
        \hline
    \end{tabular}
    \label{table:tlast_table}
\end{table}

\subsection{Analysis Based on the $T_{\rm{last}}$ Distribution}
\begin{figure*}[htb!]
\gridline {        \fig{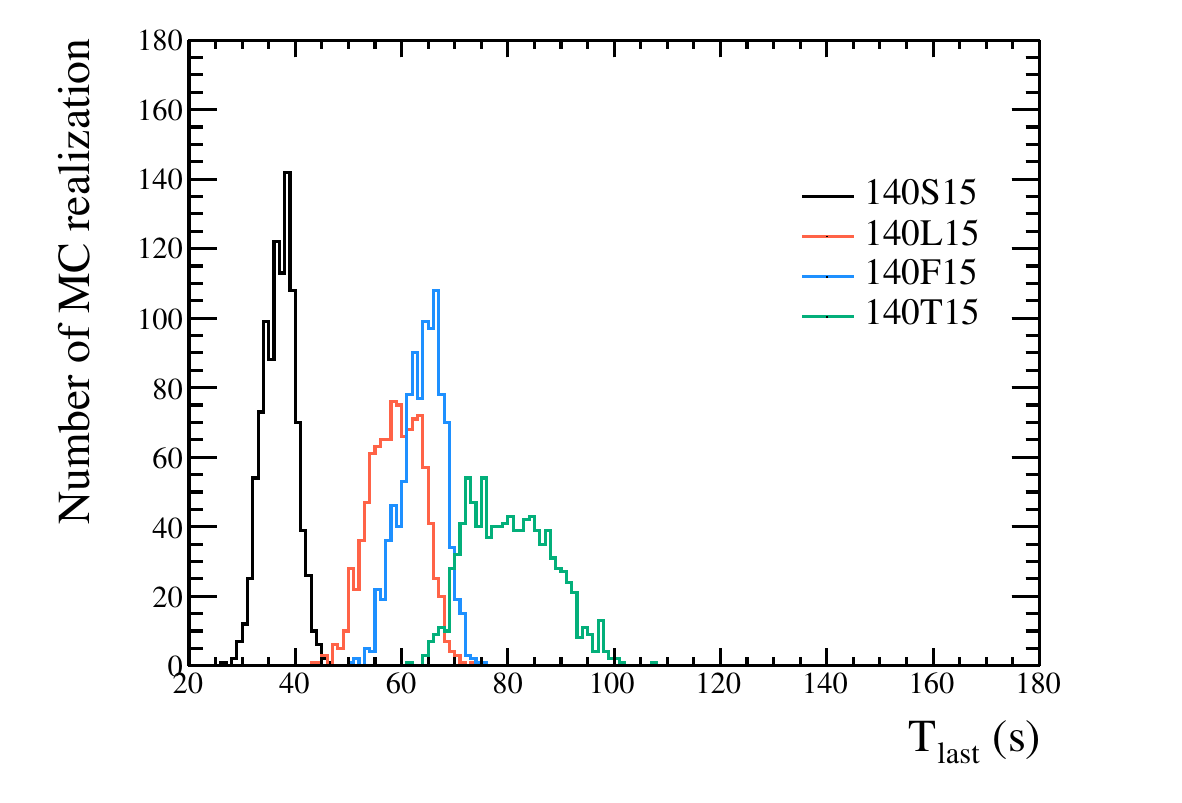}{0.35\textwidth}{(a) $M_b=1.40M_{\odot},~M_{\rm{ZAMS}}=15M_\odot$}
        \fig{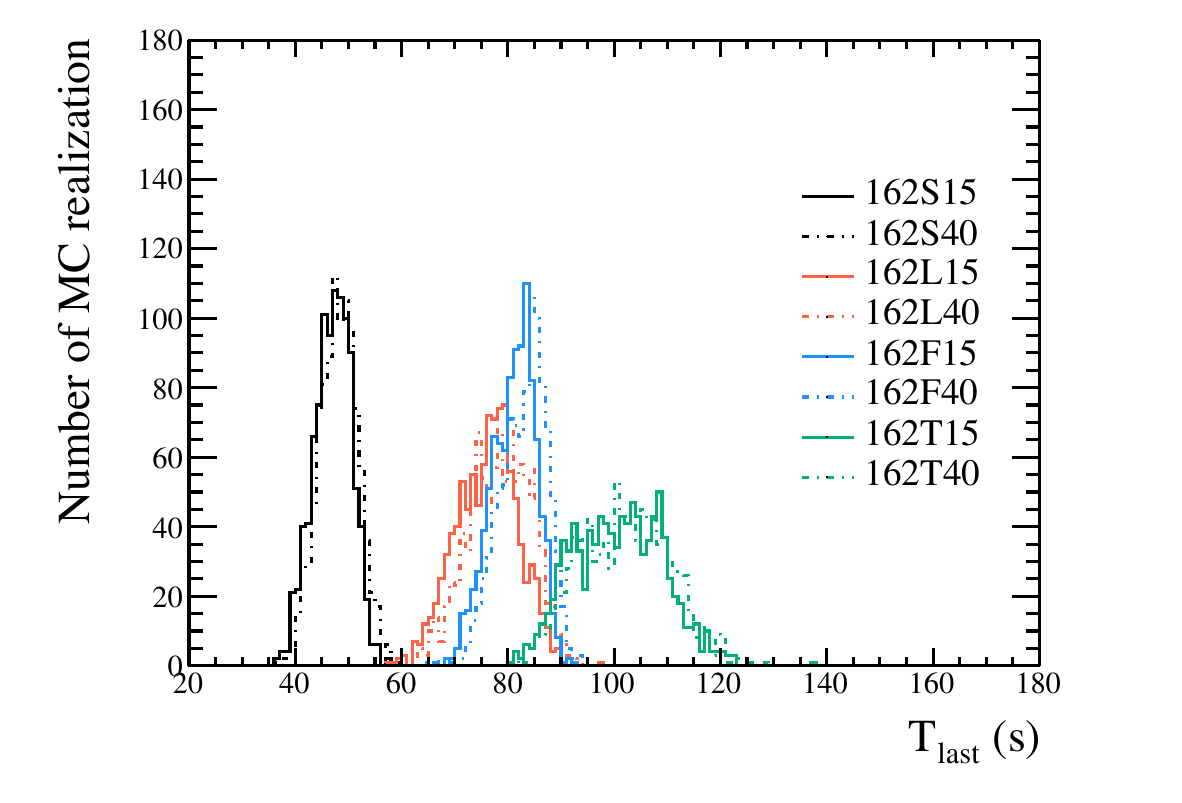}{0.35\textwidth}{(b) $M_b=1.62M_{\odot},~M_{\rm{ZAMS}}=15M_\odot,~40M_{\odot}$}
        \fig{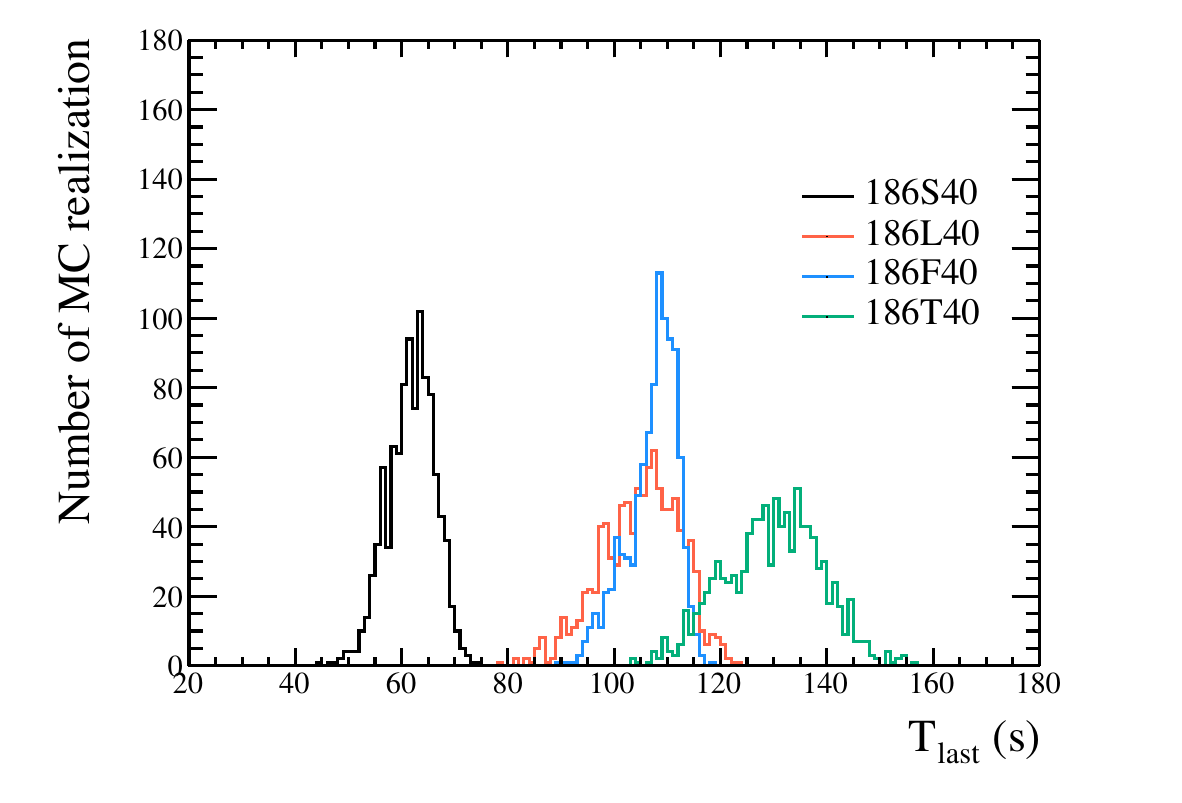}{0.35\textwidth}{(c) $M_b=1.86M_{\odot},~M_{\rm{ZAMS}}=40M_\odot$}
}
    \caption{$T_{\rm{last}}$ distribution. The black, red, blue, and green lines represent Shen EOS, LS220 EOS, Furusawa-Togashi EOS, and Togashi EOS, respectively. The horizontal axis shows $T_{\rm{last}}$ for each MC realization and the vertical axis shows the number of MC realizations in 1 sec bins. The solid and grid lines in panel (b) mean a $M_{\rm{ZAMS}} = 15M_{\odot}$ and $M_{\rm{ZAMS}} = 40M_{\odot}$, respectively.}
    \label{fig:time_last}
\end{figure*}

\begin{figure}[ht]
   \begin{center}
      \includegraphics[scale = 0.4]{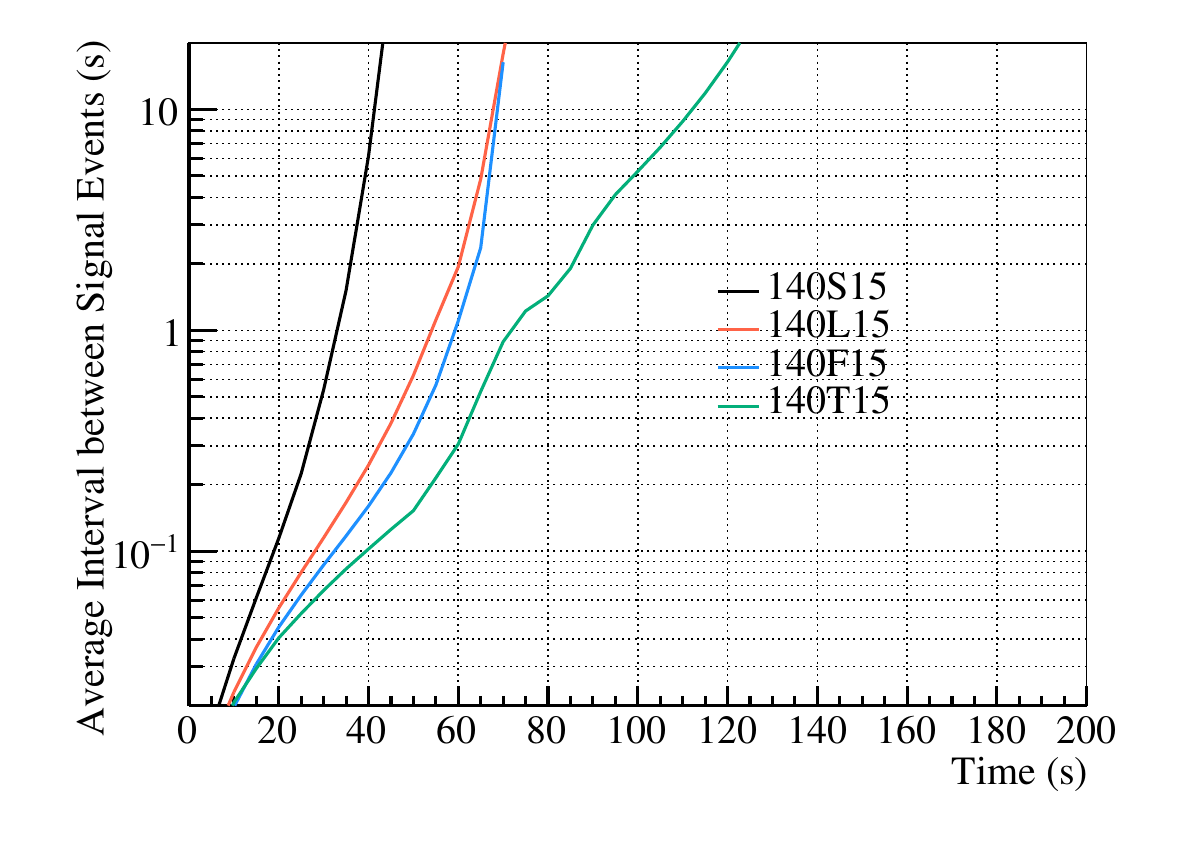}
      \caption{Average interval between signal events as a function of time after the bounce. The black, red, blue, and green represent Shen EOS, LS220 EOS, Furusawa-Togashi EOS, and Togashi EOS, respectively.}
   \label{fig:interval}
   \end{center}
\end{figure}

The $T_{\rm{last}}$ distributions for several models and for different PNS baryonic masses as determined by the method described in Section~\ref{cap:technique} are shown in Figure \ref{fig:time_last}. 
Summary information from those distributions is additionally presented in Table~\ref{table:tlast_table}. 
Comparing Figure~\ref{fig:Tlast_true_140} and Figure~\ref{fig:time_last} (a) shows that the value of $T_{\rm{last}}$ tends to be smaller than $T^{\rm{true}}_{\rm{last}}$. 
This is because the method using $T_{\rm{wid}}=5~\rm{sec}$ is more likely to select events earlier than 
$T^{\rm{true}}_{\rm{last}}$ in the final stage of PNS cooling where the interval between signal events is expected to be only a few seconds, and also because some events are lost due to the energy threshold cut.
Figure~\ref{fig:interval} shows the time evolution of signal intervals, which increase beyond 5 seconds as the final stage of PNS cooling.
In the present analysis, the last observed event is required to be identified as a signal with a significance exceeding $5\sigma$, which leads to $T_{\rm{last}}$ rarely coinciding with $T^{\rm{true}}_{\rm{last}}$.
Nevertheless, as shown in Figure~\ref{fig:time_last}, the expected differences among models can be seen.
Consistent with the trend shown in Figure~\ref{fig:Tlast_true_140}, the Shen EOS results in the shortest $T_{\rm{last}}$, while the Togashi EOS shows the longest.
Furthermore, $T_{\rm{last}}$ tends to become longer with the PNS baryonic mass.

\begin{figure*}[htb!]
\gridline {
        \fig{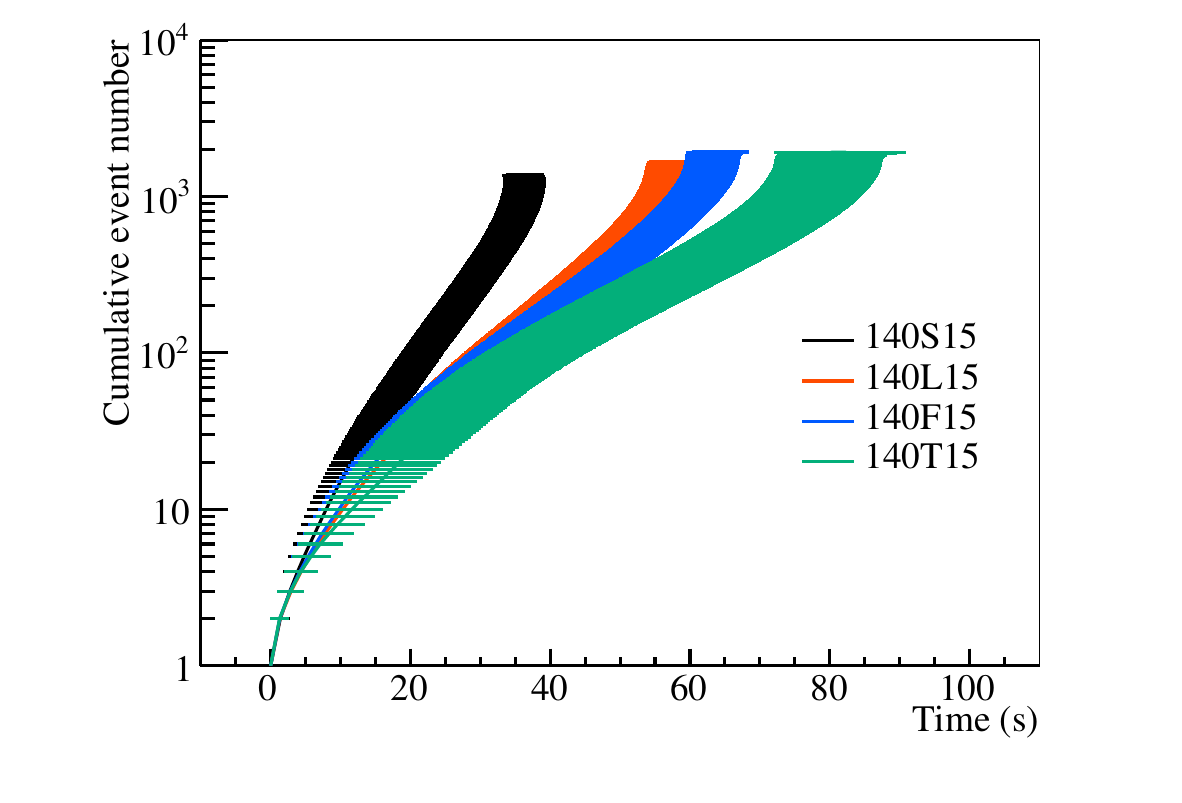}{0.53\textwidth}{(a) 
        Error bars evaluated from $T_{\rm{last}}$.
        }
        \fig{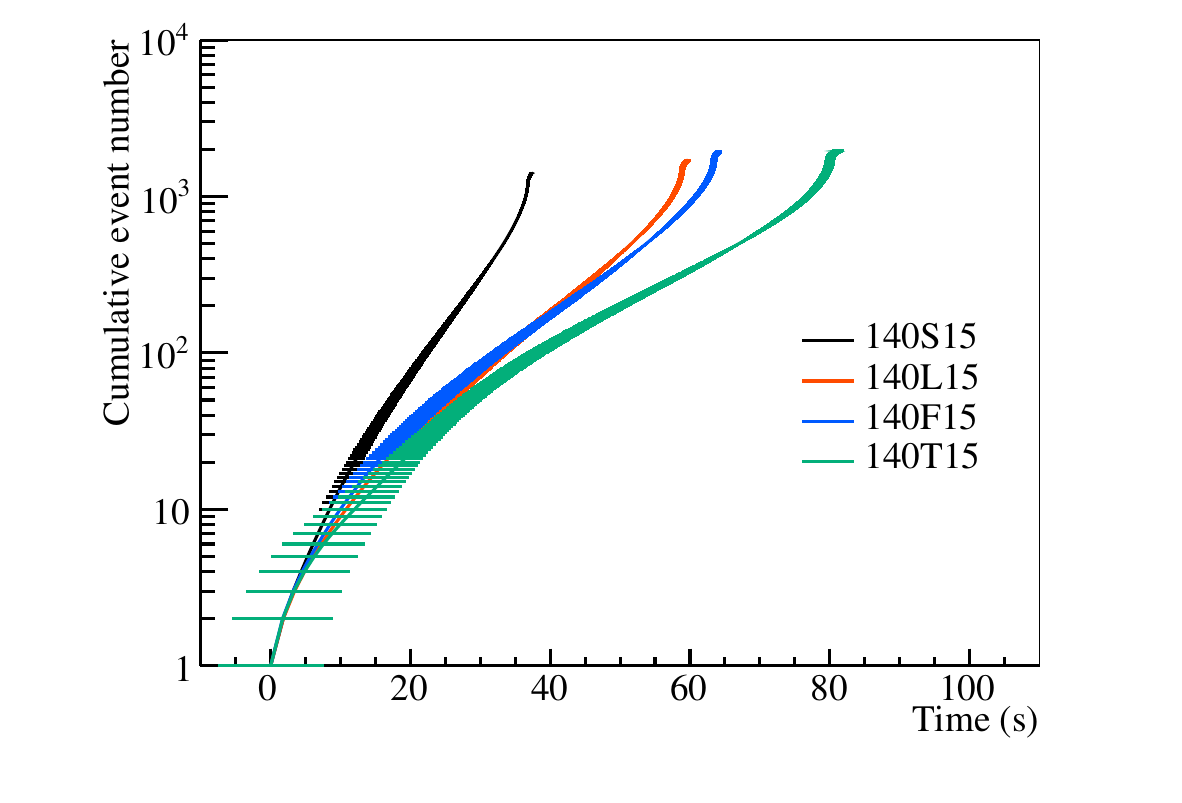}{0.53\textwidth}{(b) 
        Error bars evaluated from $T_{500}$.
        }
}
    \caption{Backward time analysis comparing different EOSs. The horizontal axis represents backward time, which sets the last observed event as the time origin, and the vertical axis represents the cumulative event number from that event.  The black, red, blue, and green represent Shen EOS, LS220 EOS, Furusawa-Togashi EOS, and Togashi EOS, respectively.
    In both panels, the curves show the expected time evolution of the cumulative event number starting from $T_{\rm{last}}$. In panel (a), the timing uncertainties are evaluated with the time origin set at $T_{\rm{last}}$ and are indicated by the error bars.
    In panel (b), the time origin is shifted to the $500\text{-th}$-to-last event ($T_{\rm{500}}$), and the error bars reflect the reduced timing uncertainties achieved by referencing this earlier event.
    }
    \label{fig:backward}
\end{figure*}

We use the cumulative event number to estimate how long neutrinos are detectable.
Figure~\ref{fig:backward} (a) shows the backward time distribution for the model in Figure~\ref{fig:time_last} (a).
The curves show the expected cumulative event distributions, plotted backward in time from $T_{\rm{last}}$, which is taken as the time origin.
Horizontal error bars indicate the range of times falling within $1\sigma$ around the mean estimated from 1000 MC simulations.
The backward cumulative distribution exhibits a different time profile depending on the EOS model and PNS mass.
In particular, the Shen EOS and the Togashi EOS distributions show a relatively large separation, while LS220 and Furusawa-Togashi EOS overlap somewhat.
These features reflect the differences in luminosity evolution as described in Section~\ref{cap:technique}.

When measured relative to $T_{\rm last}$, the increasing size of the horizontal error bars as the event rate accumulates gives the appearance 
of considerable overlap between models.
Figure~\ref{fig:backward} (b) shows the distribution when the time baseline is set to that of the $500\text{-th}$-to-last event,
while still plotting the event accumulation starting from $T_{\rm last}$.  
The curves are the same as those in Figure~\ref{fig:backward} (a), but the error bars indicate the magnitude of the time uncertainty, measured from the updated reference event.
This shift in the reference clearly reduces the width of the error bars and better indicates the differences in the early time event distributions between models.

\begin{figure*}[htb!]
\gridline {
        \fig{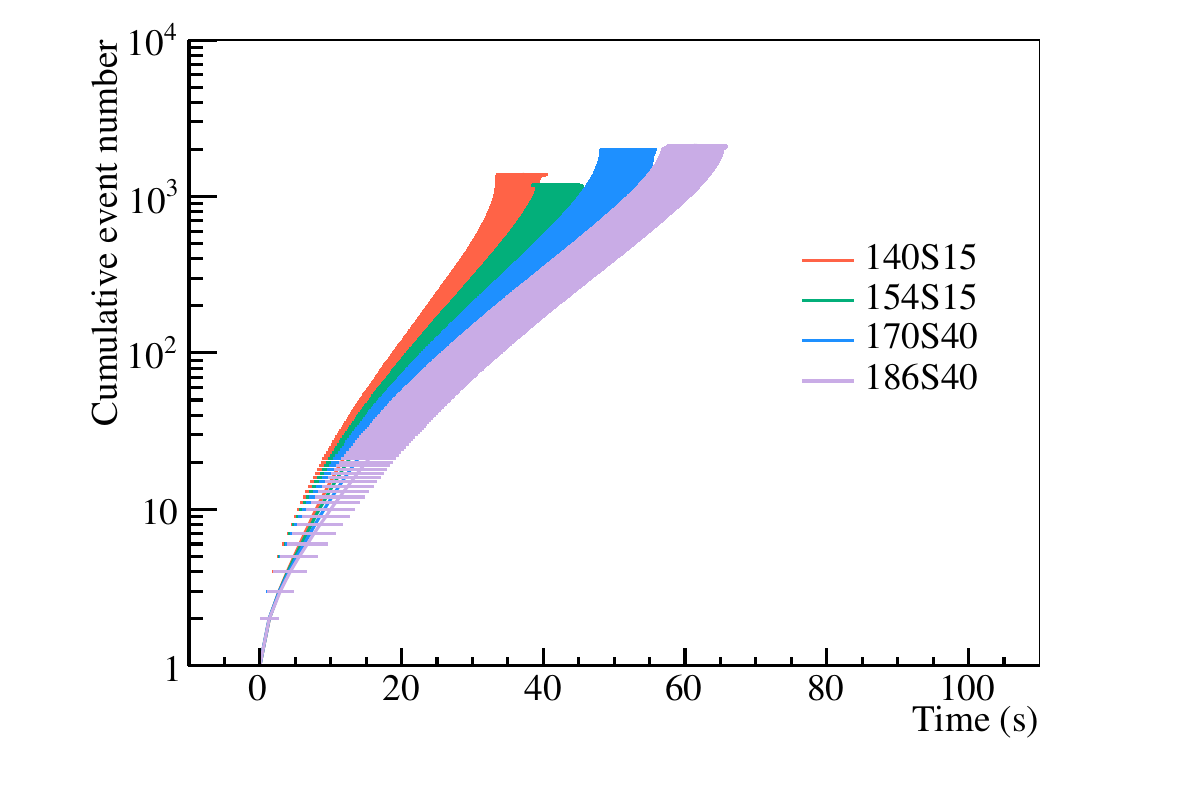}{0.53\textwidth}{(a) 
        Error bars evaluated from $T_{\rm{last}}$.
        }
        \fig{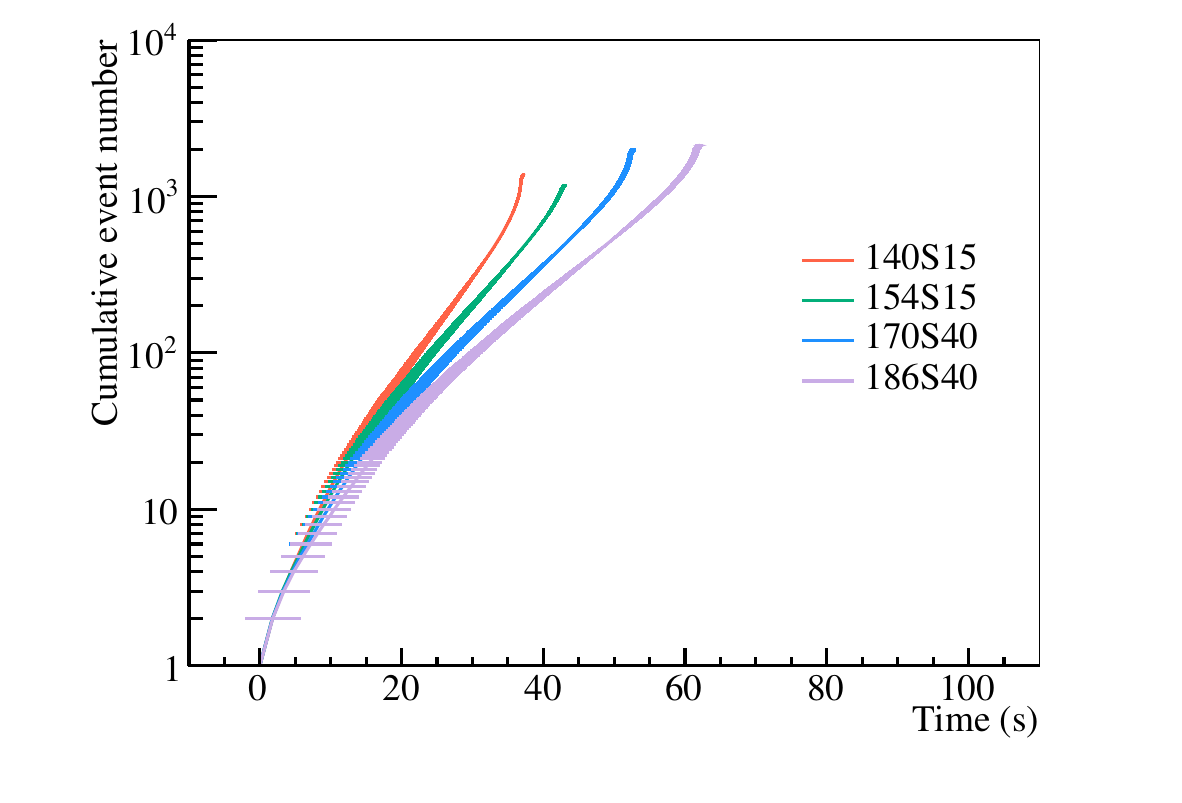}{0.53\textwidth}{(b) 
        Error bars evaluated from $T_{500}$.
        }
}
    \caption{
    Same as Figure~\ref{fig:backward}, but showing the comparison of different PNS baryonic masses in Shen EOS. The red, green, blue, and light purple represent $M_b=1.40M_{\odot}$, $M_b=1.54M_{\odot}$, $M_b=1.70M_{\odot}$, and $M_b=1.86M_{\odot}$, respectively.
    }
    \label{fig:backward_mas}
\end{figure*}

The same backward analysis applied to models with differing PNS baryonic masses is shown in Figure \ref{fig:backward_mas}. 
Both panels use the same time origins as the corresponding panel in Figure \ref{fig:backward}. 
Compared to the differences between EOS models, the variations in the baryonic mass are smaller because the behavior of the backward cumulative distribution is mainly characterized by the time evolution of the luminosity. 
As shown in Figure~\ref{fig:enelumi}, the evolution of the luminosities is similar for different PNS masses when they have a common EOS. 

\begin{figure*}[htbp]
    \centering
    \gridline {
        \fig{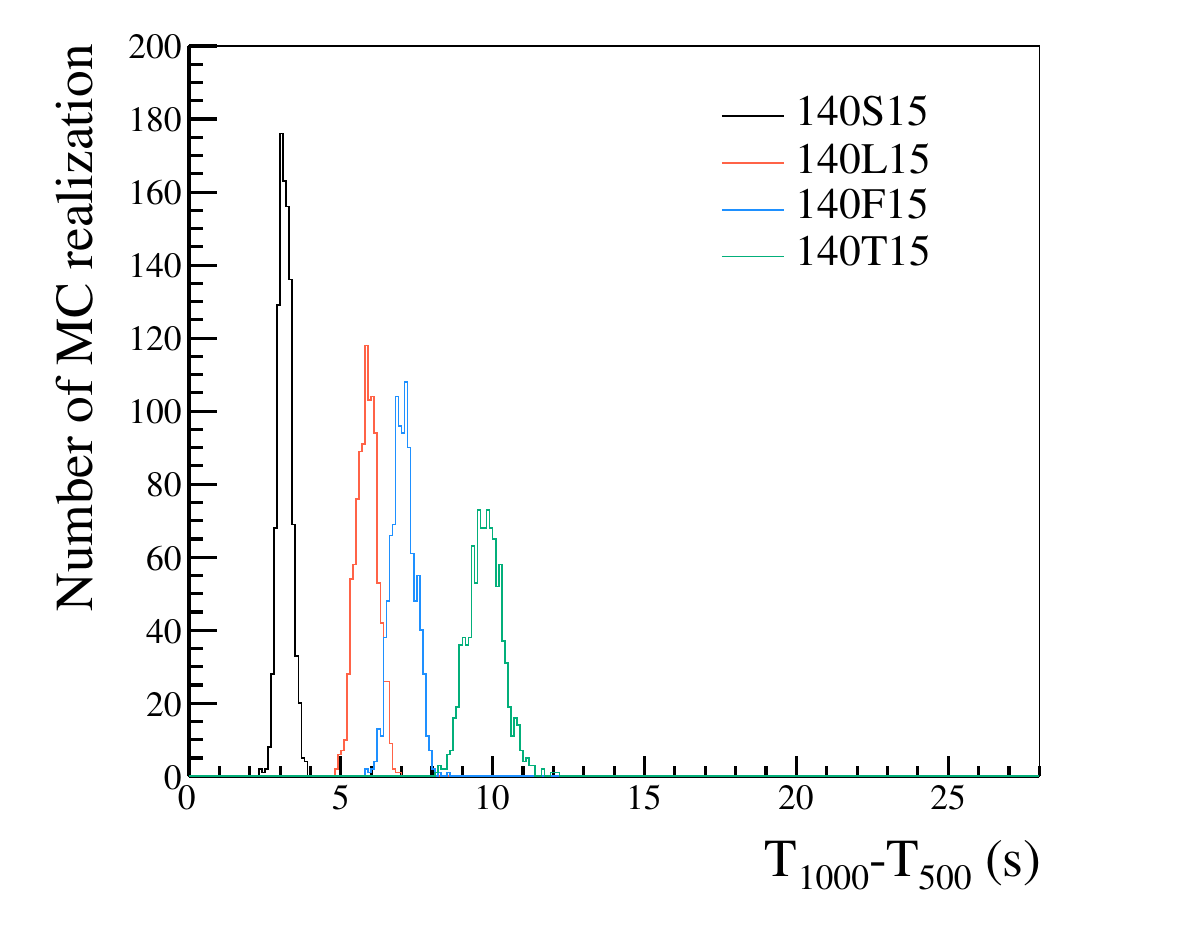}{0.35\textwidth}{(a) $M_b=1.40M_{\odot},~M_{\rm{ZAMS}}=15M_\odot$}
        \fig{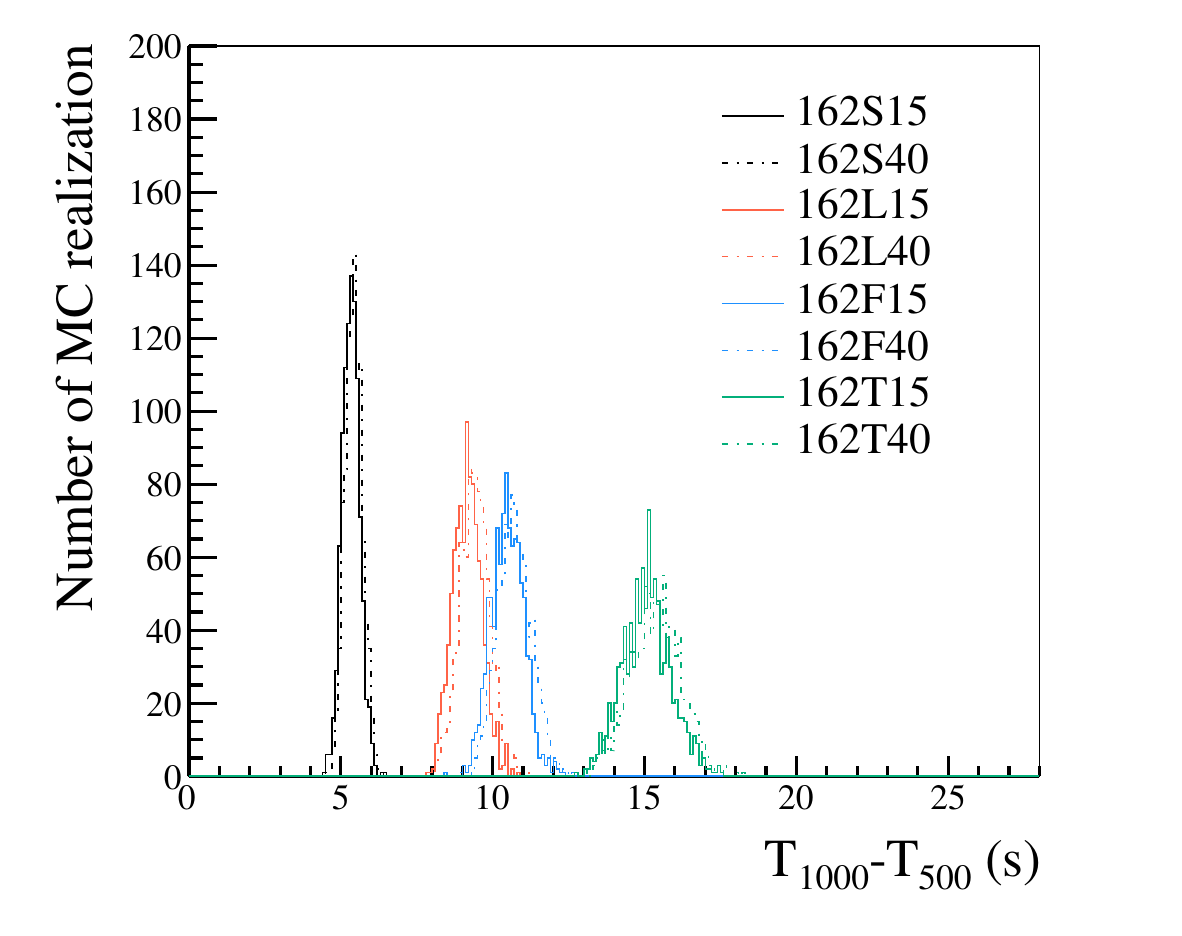}{0.35\textwidth}{(b) $M_b=1.62M_{\odot},~M_{\rm{ZAMS}}=15M_\odot,~40M_{\odot}$}
        \fig{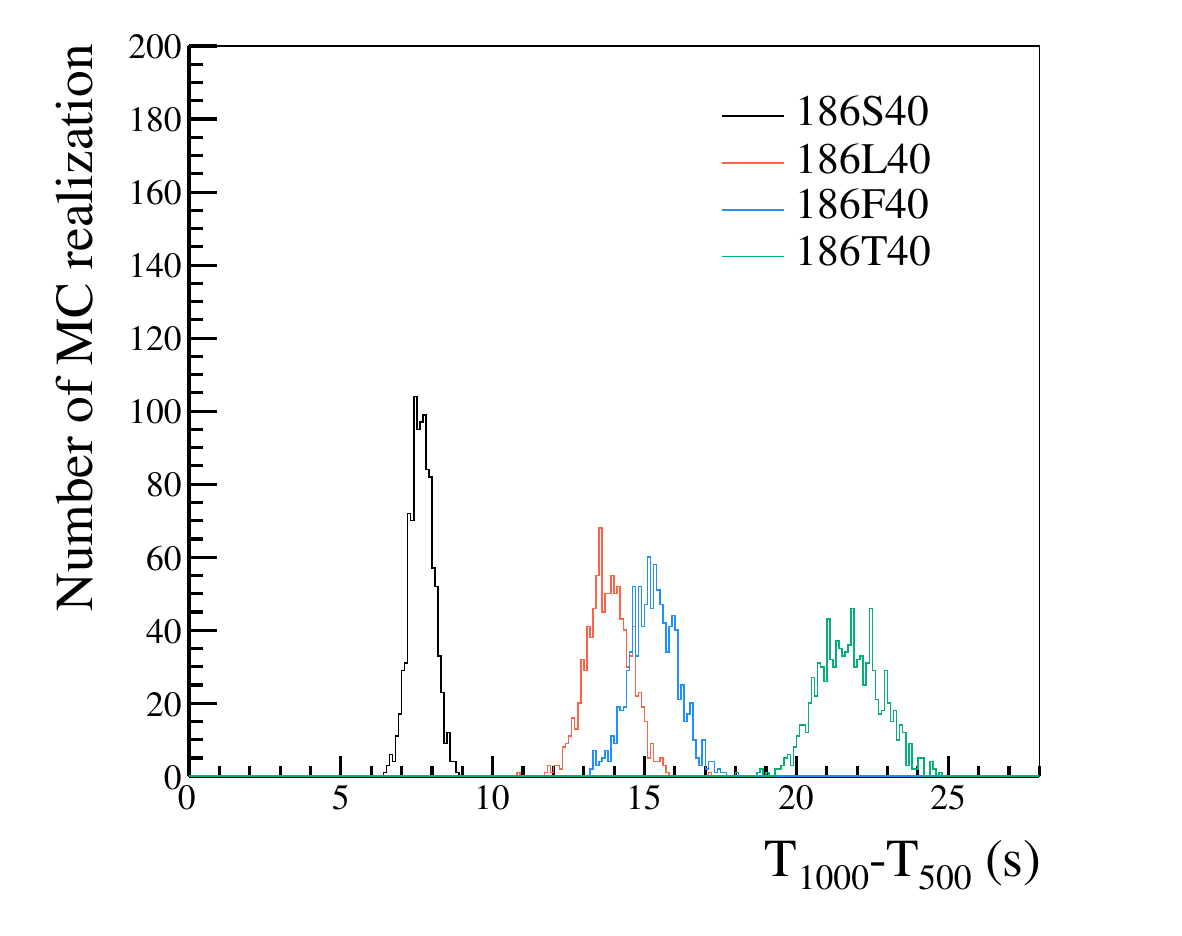}{0.35\textwidth}{(c) $M_b=1.86M_{\odot},~M_{\rm{ZAMS}}=40M_\odot$}
}
    \caption{The time difference distribution between the $500\text{-th}$-to-last event and $1000\text{-th}$-to-last event. 
    The black, red, blue, and green lines represent Shen, LS220, Furusawa-Togashi, and Togashi EOS, respectively. The horizontal axis shows the difference between time of the $500\text{-th}$-to-last event ($T_{500}$) and the time of the $1000\text{-th}$-to-last event ($T_{1000}$). 
    The solid and grid lines in panel (b) mean a $M_{\rm{ZAMS}} = 15M_{\odot}$ and $M_{\rm{ZAMS}} = 40M_{\odot}$, respectively. 
    }
    \label{fig:tdiff_500}
\end{figure*}

Analyzing intermediate events rather than only the final event improves the statistics and makes the differences between models more visible. 
Focusing on the time evolution beyond the $500\text{-th}$-to-last event counting from the last observed event, we summarize the time difference between the $500\text{-th}$-to-last and $1000\text{-th}$-to-last events in Figure~\ref{fig:tdiff_500} and Table~\ref{table:tlast_table}. 
Panel (b) of the figure compares distributes for $15$ and $40M_{\rm{ZAMS}}$ for models with the same EOS and PNS mass ($M_b = 1.62M_{\odot}$).
The differences are small and within errors as shown in the table. 
Here, the differing $M_{\rm{ZAMS}}$ values imply differences in the initial entropy. 
While analysis results using all events throughout the entire neutrino emission period do depend on the initial entropy, as discussed in \citetalias{Suwa_2019}, this time analysis using the late-phase does not depend on the physics during the early phase of collapse. 
Furthermore, some models with different combinations of PNS mass and EOS exhibit similar time differences, such as the models with 147L15 ($6.93 \pm 0.40~\rm{sec}$) and 178S40 ($6.91 \pm 0.35~\rm{sec}$) as listed in the table. 
The combination of $T_{\rm{last}}$ distribution shown in Figure~\ref{fig:time_last} and the backward time analysis presented in Figure~\ref{fig:backward} and~\ref{fig:backward_mas} improves the ability to distinguish between such models.
\subsection{Demonstration of Model Discrimination}
\begin{figure*}[ht]
   \begin{center}
      \includegraphics[scale = 0.27]{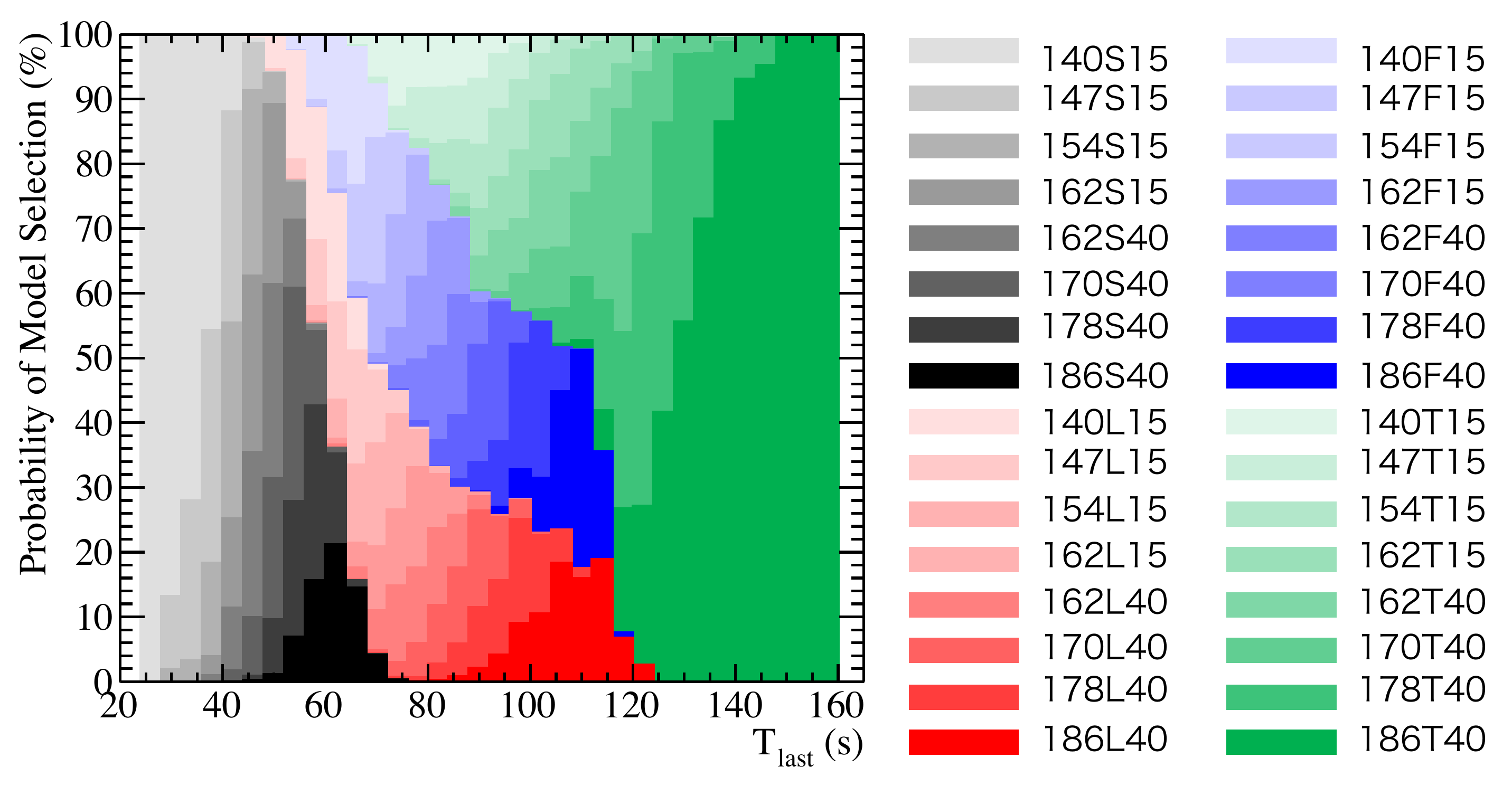}
      \caption{The probability of model selection for each model (vertical axis) given a certain $T_{\rm{last}}$ (horizontal axis), with a bin width of 4 sec. The color gradients represent different models: Shen (gray to black), LS220 (light red to red), Furusawa-Togashi (light blue to blue), and Togashi (light green to green). The differences in color intensity represent variations in the baryonic mass of the PNS, with darker colors indicating higher mass.}
   \label{fig:bayes_tlast}
   \end{center}
\end{figure*}

Here we investigate the potential of using $T_{\rm{last}}$ for supernova model discrimination in the presence of experimental background.
We adopt a Bayesian framework, where the posterior probability of a certain model is given by Bayes' theorem:
\begin{align}
    p(\text{model} | T_{\rm{last}}) &= 
    \frac{{\rm{PDF}}(T_{\rm{last}} |  \text{model}) \times P(\text{model})}
    { \sum\limits_{\text{model}} 
    \left[ {\rm{PDF}}(T_{\rm{last}} | \text{model}) \times P(\text{model}) \right] }
    \label{equ:bayes}
\end{align}
 where ``model'' denotes those in Table~\ref{table:tlast_table}, each characterized by a different EOS, $M_b$ and $M_{\rm{ZAMS}}$. Meanwhile,  ${\rm{PDF}}(T_{\rm{last}}, \text{model})$ represents the probability density function (PDF) obtained from Figure~\ref{fig:time_last} and $P(\text{model})$ is the prior probability. 
 In this analysis, we assume a uniform prior, assigning an equal probability to all 32 models, i.e., $P(\text{model})= 1/32$. 
 Here, models of $M_b=1.62M_{\odot}$ with different $M_{\rm{ZAMS}}$ are treated separately. 
 We evaluate $p(\text{model} | T_{\rm{last}})$ every $4~\rm{sec}$ for $0 < T_{\rm{last}} < 160~\rm{sec}$ and present the results in Figure~\ref{fig:bayes_tlast}.
 While identifying both the PNS baryonic mass and the EOS is difficult, this analysis demonstrates that it is possible to distinguish between the 
 Shen and Togashi EOS.
 As shown in Figure~1 of \citetalias{Nakazato_2022}, the mass-radius relations of cold neutron stars differ between EOS models. 
 For a given neutron star mass, the Shen EOS corresponds to EOS with a large radius, and EOS models with a larger radius tend to be associated with a shorter $T_{\rm{last}}$. 
 On the other hand, the distinctive feature of the Togashi EOS is its high abundance of heavy nuclei within the PNS. 
 The nuclear composition influences $T_{\rm{last}}$, leading to longer times when more nuclei with higher mass numbers are present.
In summary, based on this probability calculation, a shorter $T_{\rm{last}}$ suggests a lower central density in the neutron star, whereas a longer $T_{\rm{last}}$ suggests the presence of heavy nuclei near the surface region.

\begin{figure*}[ht]
   \begin{center}
      \includegraphics[scale = 0.27]{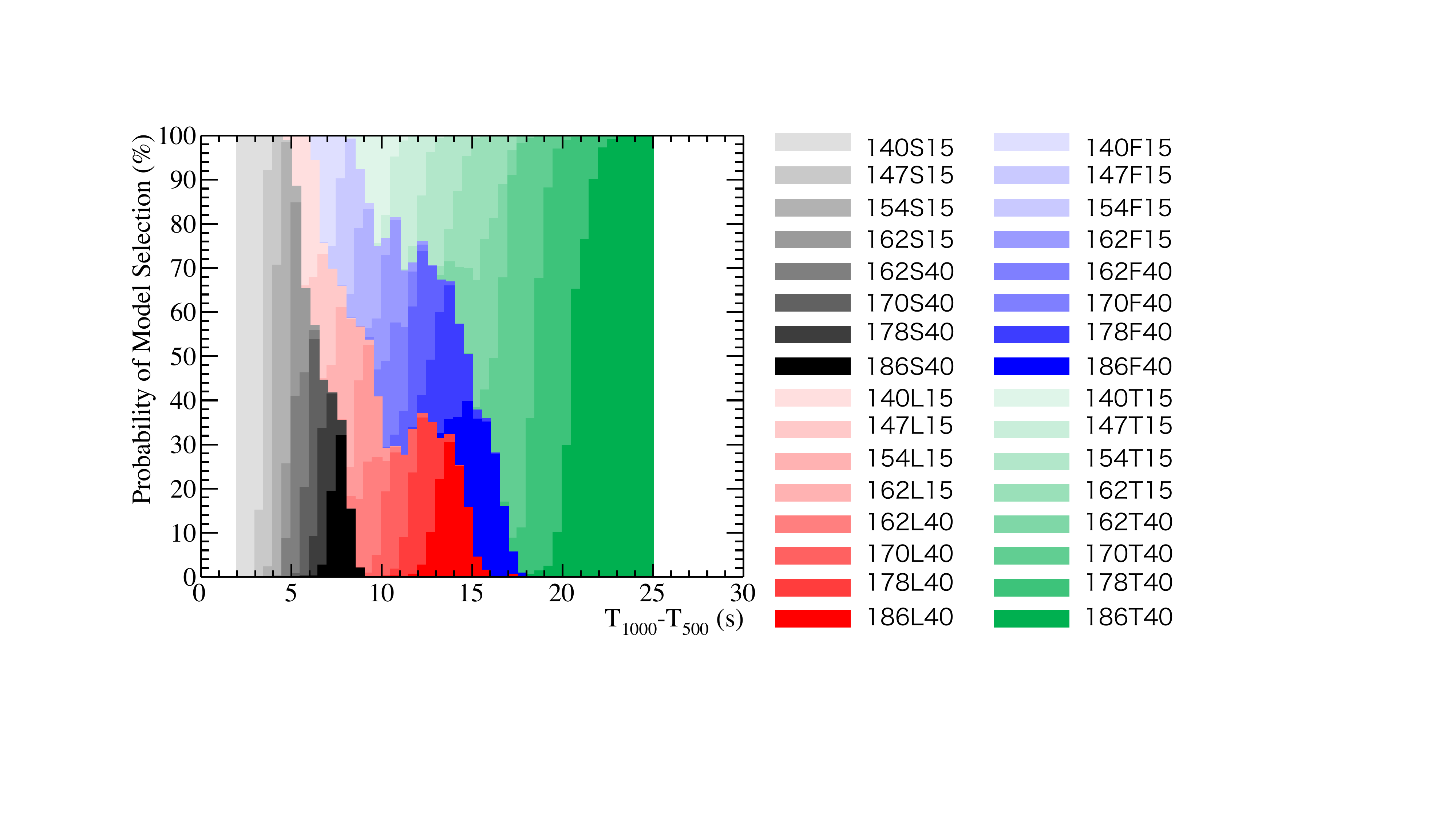}
      \caption{The probability of model selection (vertical axis) given a certain time difference (horizontal axis), with a bin width of 0.5 sec. 
      The color gradients are the same as Figure~\ref{fig:bayes_tlast}.
      }
   \label{fig:bayes_tdiff}
   \end{center}
\end{figure*}

Figure~\ref{fig:bayes_tdiff} shows the model selection probability calculated using Equation~\ref{equ:bayes}, but with $T_{\rm{last}}$ changed to the time difference between the $500\text{-th}$-to-last and $1000\text{-th}$-to-last events. 
As in Figure~\ref{fig:bayes_tlast}, calculations using the time difference can also narrow down possible EOS candidates.
The impact of background contamination in both Figures~\ref{fig:bayes_tlast} and \ref{fig:bayes_tdiff} is negligible, as confirmed by mock samples without background that produce consistent results.
This negligible impact is due to the background suppression to the $5\sigma$ significance level, as discussed in Section~\ref{cap:technique}.

\section{Conclusion}

\label{cap:conclusion}
We simulated late-phase supernova neutrino events in the Super-Kamiokande detector and developed a  
new analytical framework to determine the last observed event and its time, $T_{\rm{last}}$, explicitly incorporating realistic background contamination.
We demonstrated that $T_{\rm{last}}$ can be used to differentiate between CCSNe models with different PNS masses and EOS. 
Furthermore, by performing a backward time analysis with $T_{\rm{last}}$ as the time origin, we have shown, in particular, that the Shen EOS and Togashi EOS can be effectively distinguished. 
In addition, we showed that timing information from earlier events, such as  $T_{\rm{500}}$, can reduce statistical fluctuations and improve discrimination
between models with similar time evolution like those with the LS220 EOS and Furusawa-Togashi EOS.
However, we also found that changing $M_{\rm{ZAMS}}$ in models with the same EOS did not appreciably change the available information from the late phase neutrino emission.
This suggests that, unlike the early phase neutrino emission, late-phase neutrino emission does not depend on complicated physics, but depends more on simple parameters such as the radius and mass of the PNS (c.f.~\citetalias{Suwa_2019}). 
Finally, using a simple Bayesian calculation we demonstrated that $T_S{\rm{last}}$ can be used to constrain EOS models though it is challenging to 
also simultaneously constrain the PNS baryonic mass. 
Specifically, a shorter $T_{\rm{last}}$ suggests a lower central density in the neutron star, whereas a longer $T_{\rm{last}}$ indicates the presence of heavy nuclei near the surface region.

When the next supernova neutrino is observed, our analysis method will enable us to extract characteristics of the PNS and the properties of nuclear matter in the core.  
Moreover, it will be a useful addition to more detailed analyses incorporating more than event timing information.
Though this paper has presented our method in the contect of the Super-Kamiokande detector, it can be similarly applied to 
any other neutrino detector with event-by-event timing information.

\section*{Acknowledgment}
This work is supported by JSPS KAKENHI Grant Numbers JP19H05811, JP20H00174, JP20H01904, JP20H01905 JP20K03973,  JP20H04747, JP21K13913, JP23KJ2150, JP23KJ1609, JP24H02236, JP24H02245, JP24K00632, JP24K00668, JP24K07021, JP25K01035, and JP25H01273.
%
Numerical computations in this study were partially performed on the computing system at Osaka University's Research Center for Nuclear Physics (RCNP).
This work is supported in part by the Inter-University Research Program of the Institute for Cosmic Ray Research (ICRR), the University of Tokyo.

\clearpage
\appendix
\section{Expected Event for Each Interaction Case in SK}
In Table~\ref{table:model}, we summarize the expected number of events for each interaction case, calculated assuming the full-volume observation in SK.
\begin{table*}[hbtb]
    \caption{Expected number of events for each interaction channel for the PNS cooling models. These event numbers are calculated under the assumption that the full volume of SK is used. The total number of events includes all neutrinos with energies above $0~\rm{MeV}$.}
    \label{table:model}
    \centering
    \begin{tabular}{cccc|ccccccc}
         \hline
         Model & $M_{\rm{ZAMS}}(M_\odot)$ & EOS & $M_b(M_\odot)$ & Total & IBD & ES & CC($\nu_e,e^-$) & CC($\Bar{\nu}_e,e^+$) & NC($\nu,p$) & NC($\nu,n$) \\
         \hline \hline
         140S15 & 15 & Shen & 1.40 & 2836.8 & 2538.2 & 264.8 & 2.0 & 11.0 & 16.5 & 4.4 \\  
         147S15 & 15 & Shen & 1.47 & 2549.9 & 2274.8 & 246.5 & 1.7 & 9.9 & 13.3 & 3.6 \\
         154S15 & 15 & Shen & 1.54 & 2486.3 & 2217.6 & 243.2 & 1.7 & 9.5 & 11.2 & 3.0 \\
         162S15 & 15 & Shen & 1.62 & 2413.4 & 2151.4 & 238.7 & 2.1 & 9.2 & 9.5 & 2.6 \\
         162S40 & 40 & Shen & 1.62 & 4048.6 & 3634.0 & 367.1 & 2.8 & 15.6 & 22.9 & 6.2 \\
         170S40 & 40 & Shen & 1.70 & 4179.7 & 3748.1 & 382.0 & 3.2 & 16.6 & 23.4 & 6.3 \\
         178S40 & 40 & Shen & 1.78 & 4289.1 & 3842.6 & 396.1 & 3.4 & 17.2 & 23.3 & 6.3 \\
         186S40 & 40 & Shen & 1.86 & 4421.5 & 3958.8 & 412.0 & 3.7 & 17.8 & 23.0 & 6.2\\
         
         \hline
         140L15 & 15 & LS220 & 1.40 & 3413.2 & 3059.7 & 307.9 & 3.5 & 15.9 & 20.6 & 5.6 \\  
         147L15 & 15 & LS220 & 1.47 & 3185.4 & 2846.9 & 298.4 & 3.3 & 14.6 & 17.4 & 4.7 \\
         154L15 & 15 & LS220 & 1.54 & 3070.8 & 2740.7 & 295.0 & 3.4 & 13.5 & 14.4 & 3.9 \\
         162L15 & 15 & LS220 & 1.62 & 2919.1 & 2600.8 & 288.3 & 3.3 & 12.2 & 11.4 & 3.1 \\
         162L40 & 40 & LS220 & 1.62 & 4855.6 & 4359.8 & 430.3 & 5.4 & 23.0 & 29.2 & 7.9 \\
         170L40 & 40 & LS220 & 1.70 & 5001.0 & 4483.1 & 449.2 & 6.3 & 24.3 & 29.9 & 8.1 \\
         178L40 & 40 & LS220 & 1.78 & 5124.7 & 4586.9 & 467.9 & 7.3 & 25.2 & 29.5 & 8.0 \\
         186L40 & 40 & LS220 & 1.86 & 5256.6 & 4698.4 & 487.7 & 8.2 & 25.8 & 28.7 & 7.8 \\
         \hline
         140F15 & 15 & Furusawa-Togashi & 1.40 & 3958.9 & 3551.7 & 358.4 & 4.2 & 18.1 & 20.8 & 5.6 \\  
         147F15 & 15 & Furusawa-Togashi & 1.47 & 3692.8 & 3303.2 & 345.6 & 4.2 & 17.4 & 17.6 & 4.8 \\
         154F15 & 15 & Furusawa-Togashi & 1.54 & 3709.7 & 3314.5 & 352.7 & 4.5 & 17.4 & 16.3 & 4.4 \\
         162F15 & 15 & Furusawa-Togashi & 1.62 & 3645.3 & 3249.8 & 355.1 & 4.7 & 16.6 & 15.0 & 4.1 \\
         162F40 & 40 & Furusawa-Togashi & 1.62 & 5648.0 & 5071.0 & 506.2 & 6.4 & 25.8 & 30.3 & 8.2 \\
         170F40 & 40 & Furusawa-Togashi & 1.70 & 6012.1 & 5395.1 & 540.3 & 7.4 & 28.4 & 32.2 & 8.7 \\
         178F40 & 40 & Furusawa-Togashi & 1.78 & 6240.1 & 5593.2 & 566.3 & 8.3 & 30.3 & 33.0 & 8.9  \\
         186F40 & 40 & Furusawa-Togashi & 1.86 & 6442.9 & 5767.8 & 592.1 & 9.2 & 31.8 & 33.1 & 8.9 \\
         \hline
         140T15 & 15 & Togashi & 1.40 & 3936.7 & 3535.7 & 350.7 & 3.4 & 16.8 & 23.6 & 6.4 \\  
         147T15 & 15 & Togashi & 1.47 & 3716.1 & 3330.5 & 338.8 & 3.5 & 16.3 & 21.3 & 5.8 \\
         154T15 & 15 & Togashi & 1.54 & 3777.6 & 3385.1 & 346.1 & 3.8 & 16.6 & 20.5 & 5.5 \\
         162T15 & 15 & Togashi & 1.62 & 3761.2 & 3367.4 & 348.5 & 4.0 & 16.4 & 19.6 & 5.3 \\
         162T40 & 40 & Togashi & 1.62 & 5634.4 & 5065.5 & 495.7 & 5.3 & 24.1 & 34.5 & 9.3 \\
         170T40 & 40 & Togashi & 1.70 & 6016.3 & 5407.3 & 529.1 & 6.2 & 26.5 & 37.1 & 10.0 \\
         178T40 & 40 & Togashi & 1.78 & 6277.2 & 5637.0 & 555.4 & 7.1 & 28.5 & 38.8 & 10.5 \\
         186T40 & 40 & Togashi & 1.86 & 6514.7 & 5845.3 & 581.0 & 7.9 & 30.1 & 39.6 & 10.7 \\
         \hline

    \end{tabular}
\end{table*}

\clearpage
\bibliography{journal}{}

\begin{thebibliography}{}
\expandafter\ifx\csname natexlab\endcsname\relax\def\natexlab#1{#1}\fi
\providecommand{\url}[1]{\href{#1}{#1}}
\providecommand{\dodoi}[1]{doi:~\href{http://doi.org/#1}{\nolinkurl{#1}}}
\providecommand{\doeprint}[1]{\href{http://ascl.net/#1}{\nolinkurl{http://ascl.net/#1}}}
\providecommand{\doarXiv}[1]{\href{https://arxiv.org/abs/#1}{\nolinkurl{https://arxiv.org/abs/#1}}}

\bibitem[{{Abe} {et~al.}(2018){Abe}, {Abe}, {Aihara}, {Aimi}, {Akutsu},
  {Andreopoulos}, {Anghel}, {Anthony}, {Antonova}, {Ashida}, {Aushev}, {Barbi},
  {Barker}, {Barr}, {Beltrame}, {Berardi}, {Bergevin}, {Berkman}, {Berns},
  {Berry}, {Bhadra}, {Bravo-Bergu{\~n}o}, {Blaszczyk}, {Blondel}, {Bolognesi},
  {Boyd}, {Bravar}, {Bronner}, {Buizza Avanzini}, {Cafagna}, {Cole}, {Calland},
  {Cao}, {Cartwright}, {Catanesi}, {Checchia}, {Chen-Wishart}, {Choi}, {Choi},
  {Coleman}, {Collazuol}, {Cowan}, {Cremonesi}, {Dealtry}, {De Rosa},
  {Densham}, {Dewhurst}, {Drakopoulou}, {Di Lodovico}, {Drapier}, {Dumarchez},
  {Dunne}, {Dziewiecki}, {Emery}, {Esmaili}, {Evangelisti},
  {Fernandez-Martinez}, {Feusels}, {Finch}, {Fiorentini}, {Fiorillo}, {Fitton},
  {Frankiewicz}, {Friend}, {Fujii}, {Fukuda}, {Fukuda}, {Ganezer}, {Giganti},
  {Gonin}, {Grant}, {Gumplinger}, {Hadley}, {Hartfiel}, {Hartz}, {Hayato},
  {Hayrapetyan}, {Hill}, {Hirota}, {Horiuchi}, {Ichikawa}, {Iijima}, {Ikeda},
  {Imber}, {Inoue}, {Insler}, {Intonti}, {Ioannisian}, {Ishida}, {Ishino},
  {Ishitsuka}, {Itow}, {Iwamoto}, {Izmaylov}, {Jamieson}, {Jang}, {Jang},
  {Jeon}, {Jiang}, {Jonsson}, {Joo}, {Kaboth}, {Kachulis}, {Kajita}, {Kameda},
  {Kataoka}, {Katori}, {Kayrapetyan}, {Kearns}, {Khabibullin}, {Khotjantsev},
  {Kim}, {Kim}, {Kim}, {Kim}, {King}, {Kishimoto}, {Kobayashi}, {Koga},
  {Konaka}, {Kormos}, {Koshio}, {Korzenev}, {Kowalik}, {Kropp}, {Kudenko},
  {Kurjata}, {Kutter}, {Kuze}, {Labarga}, {Lagoda}, {Lasorak}, {Laveder},
  {Lawe}, {Learned}, {Lim}, {Lindner}, {Litchfield}, {Longhin}, {Loverre},
  {Lou}, {Ludovici}, {Ma}, {Magaletti}, {Mahn}, {Malek}, {Maret}, {Mariani},
  {Martens}, {Marti}, {Martin}, {Marzec}, {Matsuno}, {Mazzucato}, {McCarthy},
  {McCauley}, {McFarland}, {McGrew}, {Mefodiev}, {Mermod}, {Metelko},
  {Mezzetto}, {Migenda}, {Mijakowski}, {Minakata}, {Minamino}, {Mine},
  {Mineev}, {Mitra}, {Miura}, {Mochizuki}, {Monroe}, {Moon}, {Moriyama},
  {Mueller}, {Muheim}, {Murase}, {Muto}, {Nakahata}, {Nakajima}, {Nakamura},
  {Nakaya}, {Nakayama}, {Nantais}, {Needham}, {Nicholls}, {Nishimura}, {Noah},
  {Nova}, {Nowak}, {Nunokawa}, {Obayashi}, {O'Keeffe}, {Okajima}, {Okumura},
  {Onishchuk}, {O'Sullivan}, \& {O'Sullivan}}]{2018arXiv180504163H}
{Abe}, K., {Abe}, K., {Aihara}, H., {et~al.} 2018, arXiv e-prints,
  arXiv:1805.04163, \dodoi{10.48550/arXiv.1805.04163}

\bibitem[{{Abe} {et~al.}(2024){Abe}, {Bronner}, {Hayato}, {Hiraide},
  {Hosokawa}, {Ieki}, {Ikeda}, {Imaizumi}, {Iyogi}, {Kameda}, {Kanemura},
  {Kaneshima}, {Kashiwagi}, {Kataoka}, {Kato}, {Kishimoto}, {Miki}, {Mine},
  {Miura}, {Mochizuki}, {Moriyama}, {Nagao}, {Nakahata}, {Nakano}, {Nakayama},
  {Noguchi}, {Okada}, {Okamoto}, {Orii}, {Sato}, {Sekiya}, {Shiba}, {Shimizu},
  {Shiozawa}, {Sonoda}, {Suzuki}, {Takeda}, {Takemoto}, {Takenaka}, {Tanaka},
  {Watanabe}, {Yano}, {Han}, {Kajita}, {Okumura}, {Tashiro}, {Tomiya}, {Wang},
  {Wang}, {Yoshida}, {Bravo-Bergu{\~n}o}, {Fernandez}, {Labarga}, {Ospina},
  {Zaldivar}, {Pointon}, {Blaszczyk}, {Kachulis}, {Kearns}, {Raaf}, {Stone},
  {Wan}, {Wester}, {Bian}, {Griskevich}, {Kropp}, {Locke}, {Smy}, {Sobel},
  {Takhistov}, {Weatherly}, {Yankelevich}, {Ganezer}, {Hill}, {Jang}, {Kim},
  {Lee}, {Lim}, {Moon}, {Park}, {Bodur}, {Scholberg}, {Walter},
  {Beauch{\^e}ne}, {Bernard}, {Coffani}, {Drapier}, {El Hedri}, {Giampaolo},
  {Imber}, {Mueller}, {Paganini}, {Rogly}, {Quilain}, {Santos}, {Nakamura},
  {Jang}, {Machado}, {Learned}, {Matsuno}, {Iovine}, {Choi}, {Cao}, {Anthony},
  {Litchfield}, {Prouse}, {Marin}, {Scott}, {Sztuc}, {Uchida}, {Berardi},
  {Catanesi}, {Intonti}, {Radicioni}, {Calabria}, {De Rosa}, {Langella},
  {Collazuol}, {Iacob}, {Lamoureux}, {Mattiazzi}, {Ludovici}, {Gonin},
  {P{\'e}riss{\'e}}, {Pronost}, {Fujisawa}, {Maekawa}, {Nishimura}, {Okazaki},
  {Friend}, {Hasegawa}, {Ishida}, {Jakkapu}, {Kobayashi}, {Matsubara},
  {Nakadaira}, {Nakamura}, {Oyama}, {Sakashita}, {Sekiguchi}, {Tsukamoto},
  {Boschi}, {Bhuiyan}, {Burton}, {Gao}, {Goldsack}, {Katori}, {Di Lodovico},
  {Migenda}, {Molina Sedgwick}, {Ramsden}, {Taani}, {Xie}, {Zsoldos}, {Abe},
  {Hasegawa}, {Isobe}, {Kotsar}, {Miyabe}, {Ozaki}, {Shiozawa}, {Sugimoto},
  {Suzuki}, {Takagi}, {Takeuchi}, {Yamamoto}, {Zhong}, {Ashida}, {Feng},
  {Feng}, {Hayashino}, {Hirota}, {Hu}, {Hu}, {Jiang}, {Kawaue}, {Kikawa},
  {Mori}, {Nakamura}, {Nakaya}, {Wendell}, {Yasutome}, {Jenkins}, {McCauley},
  {Mehta}, {Pritchard}, {Tarrant}, {Wilking}, {Fukuda}, {Itow}, {Menjo},
  {Murase}, {Ninomiya}, {Niwa}, {Tsukada}, {Yoshioka}, {Frankiewicz}, {Lagoda},
  {Mandal}, \& {Mijakowski}}]{2024PhRvD.109i2001A}
{Abe}, K., {Bronner}, C., {Hayato}, Y., {et~al.} 2024, \prd, 109, 092001,
  \dodoi{10.1103/PhysRevD.109.092001}

\bibitem[{{Abe} {et~al.}(2022){Abe}, {Asami}, {Eizuka}, {Futagi}, {Gando},
  {Gando}, {Gima}, {Goto}, {Hachiya}, {Hata}, {Hosokawa}, {Ichimura}, {Ieki},
  {Ikeda}, {Inoue}, {Ishidoshiro}, {Kamei}, {Kawada}, {Kishimoto}, {Koga},
  {Kurasawa}, {Maemura}, {Mitsui}, {Miyake}, {Nakahata}, {Nakamura},
  {Nakamura}, {Nakamura}, {Ozaki}, {Sakai}, {Sambonsugi}, {Shimizu}, {Shirai},
  {Shiraishi}, {Suzuki}, {Suzuki}, {Takeuchi}, {Tamae}, {Watanabe}, {Yoshida},
  {Obara}, {Ichikawa}, {Yoshida}, {Umehara}, {Fushimi}, {Kotera}, {Urano},
  {Berger}, {Fujikawa}, {Learned}, {Maricic}, {Axani}, {Winslow}, {Fu},
  {Smolsky}, {Efremenko}, {Karwowski}, {Markoff}, {Tornow}, {Li}, {Detwiler},
  {Enomoto}, {Decowski}, {Grant}, {Song}, {O'Donnell}, \&
  {Dell'Oro}}]{2022ApJ...934...85A}
{Abe}, S., {Asami}, S., {Eizuka}, M., {et~al.} 2022, \apj, 934, 85,
  \dodoi{10.3847/1538-4357/ac7a3f}

\bibitem[{{Abi} {et~al.}(2020){Abi}, {Acciarri}, {Acero}, {Adamov}, {Adams},
  {Adinolfi}, {Ahmad}, {Ahmed}, {Alion}, {Alonso Monsalve}, {Alt}, {Anderson},
  {Andreopoulos}, {Andrews}, {Andrianala}, {Andringa}, {Ankowski}, {Antonova},
  {Antusch}, {Aranda-Fernandez}, {Ariga}, {Arnold}, {Arroyave}, {Asaadi},
  {Aurisano}, {Aushev}, {Autiero}, {Azfar}, {Back}, {Back}, {Backhouse},
  {Baesso}, {Bagby}, {Bajou}, {Balasubramanian}, {Baldi}, {Bambah}, {Barao},
  {Barenboim}, {Barker}, {Barkhouse}, {Barnes}, {Barr}, {Barranco Monarca},
  {Barros}, {Barrow}, {Bashyal}, {Basque}, {Bay}, {Bazo Alba}, {Beacom},
  {Bechetoille}, {Behera}, {Bellantoni}, {Bellettini}, {Bellini},
  {Beltramello}, {Belver}, {Benekos}, {Bento Neves}, {Berger}, {Berkman},
  {Bernardini}, {Berner}, {Berns}, {Bertolucci}, {Betancourt}, {Bezawada},
  {Bhattacharjee}, {Bhuyan}, {Biagi}, {Bian}, {Biassoni}, {Biery}, {Bilki},
  {Bishai}, {Bitadze}, {Blake}, {Blanco Siffert}, {Blaszczyk}, {Blazey},
  {Blucher}, {Boissevain}, {Bolognesi}, {Bolton}, {Bonesini}, {Bongrand},
  {Bonini}, {Booth}, {Booth}, {Bordoni}, {Borkum}, {Boschi}, {Bostan}, {Bour},
  {Boyd}, {Boyden}, {Bracinik}, {Braga}, {Brailsford}, {Brandt}, {Bremer},
  {Brew}, {Brianne}, {Brice}, {Brizzolari}, {Bromberg}, {Brooijmans}, {Brooke},
  {Bross}, {Brunetti}, {Buchanan}, {Budd}, {Caiulo}, {Calafiura}, {Calcutt},
  {Calin}, {Calvez}, {Calvo}, {Camilleri}, {Caminata}, {Campanelli},
  {Caratelli}, {Carini}, {Carlus}, {Carniti}, {Caro Terrazas}, {Carranza},
  {Castillo}, {Castromonte}, {Cattadori}, {Cavalier}, {Cavanna}, {Centro},
  {Cerati}, {Cervelli}, {Cervera Villanueva}, {Chalifour}, {Chang},
  {Chardonnet}, {Chatterjee}, {Chattopadhyay}, {Chaves}, {Chen}, {Chen},
  {Chen}, {Cherdack}, {Chi}, {Childress}, {Chiriacescu}, {Cho}, {Choubey},
  {Christensen}, {Christian}, {Christodoulou}, {Church}, {Clarke}, {Coan},
  {Cocco}, {Coelho}, {Conley}, {Conrad}, {Convery}, {Corwin}, {Cotte},
  {Cremaldi}, {Cremonesi}, {Crespo-Anad{\'o}n}, {Cristaldo}, {Cross}, {Cuesta},
  {Cui}, {Cussans}, {Dabrowski}, {Da Motta}, {Da Silva Peres}, {David},
  {Davies}, {Davini}, {Dawson}, {De}, {De Almeida}, {Debbins}, {De Bonis},
  {Decowski}, {De Gouvea}, {De Holanda}, {De Icaza Astiz}, {Deisting}, {De
  Jong}, {Delbart}, {Delepine}, {Delgado}, {Dell'Acqua}, {De Lurgio}, {De Mello
  Neto}, {DeMuth}, {Dennis}, {Densham}, \& {Deptuch}}]{2020JInst..15.8008A}
{Abi}, B., {Acciarri}, R., {Acero}, M.~A., {et~al.} 2020, JINST, 15, T08008,
  \dodoi{10.1088/1748-0221/15/08/T08008}

\bibitem[{{Akaho} {et~al.}(2023){Akaho}, {Harada}, {Nagakura}, {Iwakami},
  {Okawa}, {Furusawa}, {Matsufuru}, {Sumiyoshi}, \&
  {Yamada}}]{2023ApJ...944...60A}
{Akaho}, R., {Harada}, A., {Nagakura}, H., {et~al.} 2023, \apj, 944, 60,
  \dodoi{10.3847/1538-4357/acad76}

\bibitem[{{Alexeyev} {et~al.}(1988){Alexeyev}, {Alexeyeva}, {Krivosheina}, \&
  {Volchenko}}]{1988PhLB..205..209A}
{Alexeyev}, E.~N., {Alexeyeva}, L.~N., {Krivosheina}, I.~V., \& {Volchenko},
  V.~I. 1988, Phys. Lett. B, 205, 209, \dodoi{10.1016/0370-2693(88)91651-6}

\bibitem[{{An} {et~al.}(2016){An}, {An}, {An}, {Antonelli}, {Baussan},
  {Beacom}, {Bezrukov}, {Blyth}, {Brugnera}, {Buizza Avanzini}, {Busto},
  {Cabrera}, {Cai}, {Cai}, {Cammi}, {Cao}, {Cao}, {Chang}, {Chen}, {Chen},
  {Chen}, {Chiesa}, {Clemenza}, {Clerbaux}, {Conrad}, {D'Angelo}, {De Kerret},
  {Deng}, {Deng}, {Ding}, {Djurcic}, {Dornic}, {Dracos}, {Drapier}, {Dusini},
  {Dye}, {Enqvist}, {Fan}, {Fang}, {Favart}, {Ford}, {G{\"o}ger-Neff}, {Gan},
  {Garfagnini}, {Giammarchi}, {Gonchar}, {Gong}, {Gong}, {Gonin}, {Grassi},
  {Grewing}, {Guan}, {Guarino}, {Guo}, {Guo}, {Guo}, {Hagner}, {Han}, {He},
  {Heng}, {Hsiung}, {Hu}, {Hu}, {Hu}, {Huang}, {Huang}, {Huo}, {Ioannisian},
  {Jeitler}, {Ji}, {Jiang}, {Jollet}, {Kang}, {Karagounis}, {Kazarian},
  {Krumshteyn}, {Kruth}, {Kuusiniemi}, {Lachenmaier}, {Leitner}, {Li}, {Li},
  {Li}, {Li}, {Li}, {Li}, {Li}, {Li}, {Li}, {Liang}, {Lin}, {Lin}, {Lin},
  {Ling}, {Lippi}, {Liu}, {Liu}, {Liu}, {Liu}, {Liu}, {Liu}, {Liu}, {Liu},
  {Liu}, {Lombardi}, {Long}, {Lu}, {Lu}, {Lu}, {Lu}, {Lubsandorzhiev},
  {Ludhova}, {Luo}, {Lyashuk}, {M{\"o}llenberg}, {Ma}, {Mantovani}, {Mao},
  {Mari}, {McDonough}, {Meng}, {Meregaglia}, {Meroni}, {Mezzetto}, {Miramonti},
  {Mueller}, {Naumov}, {Oberauer}, {Ochoa-Ricoux}, {Olshevskiy}, {Ortica},
  {Paoloni}, {Peng}, {Peng}, {Previtali}, {Qi}, {Qian}, {Qian}, {Qian}, {Qin},
  {Raffelt}, {Ranucci}, {Ricci}, {Robens}, {Romani}, {Ruan}, {Ruan},
  {Salamanna}, {Shaevitz}, {Sinev}, {Sirignano}, {Sisti}, {Smirnov}, {Soiron},
  {Stahl}, {Stanco}, {Steinmann}, {Sun}, {Sun}, {Taichenachev}, {Tang},
  {Tkachev}, {Trzaska}, {van Waasen}, {Volpe}, {Vorobel}, {Votano}, {Wang},
  {Wang}, {Wang}, {Wang}, {Wang}, {Wang}, {Wang}, {Wang}, {Wang}, {Wang},
  {Wang}, {Wang}, {Wang}, {Wang}, {Wei}, {Wen}, {Wiebusch}, {Wonsak}, {Wu},
  {Wulz}, {Wurm}, {Xi}, {Xia}, {Xie}, {Xing}, {Xu}, {Yan}, {Yang}, {Yang},
  {Yang}, {Yang}, {Yang}, \& {Yao}}]{2016JPhG...43c0401A}
{An}, F., {An}, G., {An}, Q., {et~al.} 2016, J. Phys. G, 43, 030401,
  \dodoi{10.1088/0954-3899/43/3/030401}

\bibitem[{{Bionta} {et~al.}(1987){Bionta}, {Blewitt}, {Bratton}, {Casper},
  {Ciocio}, {Claus}, {Cortez}, {Crouch}, {Dye}, {Errede}, {Foster}, {Gajewski},
  {Ganezer}, {Goldhaber}, {Haines}, {Jones}, {Kielczewska}, {Kropp}, {Learned},
  {Losecco}, {Matthews}, {Miller}, {Mudan}, {Park}, {Price}, {Reines},
  {Schultz}, {Seidel}, {Shumard}, {Sinclair}, {Sobel}, {Stone}, {Sulak},
  {Svoboda}, {Thornton}, {van der Velde}, \& {Wuest}}]{1987PhRvL..58.1494B}
{Bionta}, R.~M., {Blewitt}, G., {Bratton}, C.~B., {et~al.} 1987, \prl, 58,
  1494, \dodoi{10.1103/PhysRevLett.58.1494}

\bibitem[{{Burrows} {et~al.}(1987){Burrows}, {Lattimer}, {Mazurek}, \&
  {Yahil}}]{1987STIN...8729393B}
{Burrows}, A., {Lattimer}, J.~M., {Mazurek}, T.~J., \& {Yahil}, A. 1987,
  {Research in astrophysics: Stellar collapse and supernovae}, Termination
  Report, 1 Aug. 1980 - 30 Nov. 1986 State Univ. of New York, Stony Brook.

\bibitem[{{Fukuda} {et~al.}(2003){Fukuda}, {Fukuda}, {Hayakawa}, {Ichihara},
  {Ishitsuka}, {Itow}, {Kajita}, {Kameda}, {Kaneyuki}, {Kasuga}, {Kobayashi},
  {Kobayashi}, {Koshio}, {Miura}, {Moriyama}, {Nakahata}, {Nakayama}, {Namba},
  {Obayashi}, {Okada}, {Oketa}, {Okumura}, {Oyabu}, {Sakurai}, {Shiozawa},
  {Suzuki}, {Takeuchi}, {Toshito}, {Totsuka}, {Yamada}, {Desai}, {Earl},
  {Hong}, {Kearns}, {Masuzawa}, {Messier}, {Stone}, {Sulak}, {Walter}, {Wang},
  {Scholberg}, {Barszczak}, {Casper}, {Liu}, {Gajewski}, {Halverson}, {Hsu},
  {Kropp}, {Mine}, {Price}, {Reines}, {Smy}, {Sobel}, {Vagins}, {Ganezer},
  {Keig}, {Ellsworth}, {Tasaka}, {Flanagan}, {Kibayashi}, {Learned}, {Matsuno},
  {Stenger}, {Hayato}, {Ishii}, {Ichikawa}, {Kanzaki}, {Kobayashi}, {Maruyama},
  {Nakamura}, {Oyama}, {Sakai}, {Sakuda}, {Sasaki}, {Echigo}, {Iwashita},
  {Kohama}, {Suzuki}, {Hasegawa}, {Inagaki}, {Kato}, {Maesaka}, {Nakaya},
  {Nishikawa}, {Yamamoto}, {Haines}, {Kim}, {Sanford}, {Svoboda}, {Blaufuss},
  {Chen}, {Conner}, {Goodman}, {Guillian}, {Sullivan}, {Turcan}, {Habig},
  {Ackerman}, {Goebel}, {Hill}, {Jung}, {Kato}, {Kerr}, {Malek}, {Martens},
  {Mauger}, {McGrew}, {Sharkey}, {Viren}, {Yanagisawa}, {Doki}, {Inaba}, {Ito},
  {Kirisawa}, {Kitaguchi}, {Mitsuda}, {Miyano}, {Saji}, {Takahata},
  {Takahashi}, {Higuchi}, {Kajiyama}, {Kusano}, {Nagashima}, {Nitta}, {Takita},
  {Yamaguchi}, {Yoshida}, {Kim}, {Kim}, {Yoo}, {Okazawa}, {Etoh}, {Fujita},
  {Gando}, {Hasegawa}, {Hasegawa}, {Hatakeyama}, {Inoue}, {Ishihara},
  {Iwamoto}, {Koga}, {Nishiyama}, {Ogawa}, {Shirai}, {Suzuki}, {Takayama},
  {Tsushima}, {Koshiba}, {Ichikawa}, {Hashimoto}, {Hatakeyama}, {Koike},
  {Horiuchi}, {Nemoto}, {Nishijima}, {Takeda}, {Fujiyasu}, {Futagami},
  {Ishino}, {Kanaya}, {Morii}, {Nishihama}, {Nishimura}, {Suzuki}, {Watanabe},
  {Kielczewska}, {Golebiewska}, {Berns}, {Boyd}, {Doyle}, {George}, {Stachyra},
  {Wai}, {Wilkes}, {Young}, {Kobayashi}, \& {Super-Kamiokande
  Collaboration}}]{2003NIMPA.501..418F}
{Fukuda}, S., {Fukuda}, Y., {Hayakawa}, T., {et~al.} 2003, Nucl. Instrum.
  Methods Phys. Res. A, 501, 418, \dodoi{10.1016/S0168-9002(03)00425-X}

\bibitem[{{Furusawa} {et~al.}(2017){Furusawa}, {Togashi}, {Nagakura},
  {Sumiyoshi}, {Yamada}, {Suzuki}, \& {Takano}}]{2017JPhG...44i4001F}
{Furusawa}, S., {Togashi}, H., {Nagakura}, H., {et~al.} 2017, J. Phys. G, 44,
  094001, \dodoi{10.1088/1361-6471/aa7f35}

\bibitem[{Harada(2023)}]{zenodo_harada}
Harada, A. 2023, SPECIAL BLEND: Supernova Parameter Estimation Code based on
  Insight on Analytic Late-time Burst Light curve at Earth Neutrino Detector,
  1.0,  Zenodo, \dodoi{10.5281/zenodo.8004041}

\bibitem[{{Harada} {et~al.}(2023){Harada}, {Suwa}, {Harada}, {Koshio}, {Mori},
  {Nakanishi}, {Nakazato}, {Sumiyoshi}, \& {Wendell}}]{2023ApJ...954...52H}
{Harada}, A., {Suwa}, Y., {Harada}, M., {et~al.} 2023, ApJ, 954, 52,
  \dodoi{10.3847/1538-4357/ace52e}

\bibitem[{{Hirata} {et~al.}(1987){Hirata}, {Kajita}, {Koshiba}, {Nakahata},
  {Oyama}, {Sato}, {Suzuki}, {Takita}, {Totsuka}, {Kifune}, {Suda},
  {Takahashi}, {Tanimori}, {Miyano}, {Yamada}, {Beier}, {Feldscher}, {Kim},
  {Mann}, {Newcomer}, {van}, {Zhang}, \& {Cortez}}]{1987PhRvL..58.1490H}
{Hirata}, K., {Kajita}, T., {Koshiba}, M., {et~al.} 1987, \prl, 58, 1490,
  \dodoi{10.1103/PhysRevLett.58.1490}

\bibitem[{{Horiuchi} \& {Kneller}(2018)}]{2018JPhG...45d3002H}
{Horiuchi}, S., \& {Kneller}, J.~P. 2018, J. phys. G, 45, 043002,
  \dodoi{10.1088/1361-6471/aaa90a}

\bibitem[{{Janka}(2017)}]{2017hsn..book.1575J}
{Janka}, H.-T. 2017, in Handbook of Supernovae, ed. A.~W. {Alsabti} \&
  P.~{Murdin}, 1575, \dodoi{10.1007/978-3-319-21846-5_4}

\bibitem[{{Kashiwagi} {et~al.}(2024){Kashiwagi}, {Abe}, {Bronner}, {Hayato},
  {Hiraide}, {Hosokawa}, {Ieki}, {Ikeda}, {Kameda}, {Kanemura}, {Kaneshima},
  {Kataoka}, {Miki}, {Mine}, {Miura}, {Moriyama}, {Nakano}, {Nakahata},
  {Nakayama}, {Noguchi}, {Sato}, {Sekiya}, {Shiba}, {Shimizu}, {Shiozawa},
  {Sonoda}, {Suzuki}, {Takeda}, {Takemoto}, {Tanaka}, {Yano}, {Han}, {Kajita},
  {Okumura}, {Tashiro}, {Tomiya}, {Wang}, {Yoshida}, {Fernandez}, {Labarga},
  {Ospina}, {Zaldivar}, {Pointon}, {Kearns}, {Raaf}, {Wan}, {Wester}, {Bian},
  {Griskevich}, {Locke}, {Smy}, {Sobel}, {Takhistov}, {Yankelevich}, {Hill},
  {Jang}, {Lee}, {Moon}, {Park}, {Bodur}, {Scholberg}, {Walter},
  {Beauch{\^e}ne}, {Drapier}, {Giampaolo}, {Mueller}, {Santos}, {Paganini},
  {Quilain}, {Rogly}, {Nakamura}, {Jang}, {Machado}, {Learned}, {Choi},
  {Iovine}, {Cao}, {Anthony}, {Martin}, {Prouse}, {Scott}, {Sztuc}, {Uchida},
  {Berardi}, {Catanesi}, {Radicioni}, {Calabria}, {Langella}, {De Rosa},
  {Collazuol}, {Iacob}, {Mattiazzi}, {Ludovici}, {Gonin}, {P{\'e}riss{\'e}},
  {Pronost}, {Fujisawa}, {Maekawa}, {Nishimura}, {Okazaki}, {Akutsu}, {Friend},
  {Hasegawa}, {Ishida}, {Kobayashi}, {Jakkapu}, {Matsubara}, {Nakadaira},
  {Nakamura}, {Oyama}, {Sakashita}, {Sekiguchi}, {Tsukamoto}, {Bhuiyan},
  {Burton}, {Di Lodovico}, {Gao}, {Goldsack}, {Katori}, {Migenda}, {Ramsden},
  {Xie}, {Zsoldos}, {Suzuki}, {Takagi}, {Takeuchi}, {Zhong}, {Feng}, {Feng},
  {Hu}, {Hu}, {Kawaue}, {Kikawa}, {Mori}, {Nakaya}, {Wendell}, {Yasutome},
  {Jenkins}, {McCauley}, {Mehta}, {Tarrant}, {Fukuda}, {Itow}, {Menjo},
  {Ninomiya}, {Yoshioka}, {Lagoda}, {Lakshmi}, {Mandal}, {Mijakowski},
  {Prabhu}, {Zalipska}, {Jia}, {Jiang}, {Jung}, {Shi}, {Wilking}, {Yanagisawa},
  {Harada}, {Hino}, {Ishino}, {Koshio}, {Nakanishi}, {Sakai}, {Tada}, {Tano},
  {Ishizuka}, {Barr}, {Barrow}, {Cook}, {Samani}, {Wark}, {Holin}, {Nova},
  {Jung}, {Yang}, {Yang}, {Yoo}, {Fannon}, {Kneale}, {Malek}, {McElwee},
  {Thiesse}, {Thompson}, {Wilson}, {Okazawa}, {Kim}, {Kwon}, {Seo}, {Yu},
  {Ichikawa}, {Nakamura}, {Tairafune}, {Nishijima}, {Eguchi}, {Nakagiri},
  {Nakajima}, {Shima}, {Taniuchi}, \& {Watanabe}}]{2024ApJ...970...93K}
{Kashiwagi}, Y., {Abe}, K., {Bronner}, C., {et~al.} 2024, \apj, 970, 93,
  \dodoi{10.3847/1538-4357/ad4d8e}

\bibitem[{{Kato} {et~al.}(2020){Kato}, {Ishidoshiro}, \&
  {Yoshida}}]{2020ARNPS..70..121K}
{Kato}, C., {Ishidoshiro}, K., \& {Yoshida}, T. 2020, Annu. Rev. Nucl. Part.
  Sci., 70, 121, \dodoi{10.1146/annurev-nucl-040620-021320}

\bibitem[{{Kotake} {et~al.}(2006){Kotake}, {Sato}, \&
  {Takahashi}}]{2006RPPh...69..971K}
{Kotake}, K., {Sato}, K., \& {Takahashi}, K. 2006, Rep. Prog. Phys., 69, 971,
  \dodoi{10.1088/0034-4885/69/4/R03}

\bibitem[{Lattimer \& {Swesty}(1991)}]{LS_1991}
Lattimer, J.~M., \& {Swesty}, D.~F. 1991, Nucl. Phys. A, 535, 331,
  \dodoi{https://doi.org/10.1016/0375-9474(91)90452-C}

\bibitem[{{Lattimer} \& {Yahil}(1989)}]{1989ApJ...340..426L}
{Lattimer}, J.~M., \& {Yahil}, A. 1989, \apj, 340, 426, \dodoi{10.1086/167404}

\bibitem[{{Li} {et~al.}(2021){Li}, {Roberts}, \&
  {Beacom}}]{2021PhRvD.103b3016L}
{Li}, S.~W., {Roberts}, L.~F., \& {Beacom}, J.~F. 2021, \prd, 103, 023016,
  \dodoi{10.1103/PhysRevD.103.023016}

\bibitem[{{Locke} {et~al.}(2024){Locke}, {Coffani}, {Abe}, {Bronner}, {Hayato},
  {Ikeda}, {Imaizumi}, {Ito}, {Kameda}, {Kataoka}, {Miura}, {Moriyama},
  {Nagao}, {Nakahata}, {Nakajima}, {Nakayama}, {Okada}, {Okamoto}, {Orii},
  {Pronost}, {Sekiya}, {Shiozawa}, {Sonoda}, {Suzuki}, {Takeda}, {Takemoto},
  {Takenaka}, {Tanaka}, {Yano}, {Hirade}, {Kanemura}, {Miki}, {Watabe}, {Han},
  {Kajita}, {Okumura}, {Tashiro}, {Xia}, {Wang}, {Megias}, {Bravo-Bergu{\~n}o},
  {Labarga}, {Marti}, {Zaldivar}, {Pointon}, {Blaszczyk}, {Kearns}, {Raaf},
  {Stone}, {Wan}, {Wester}, {Bian}, {Griskevich}, {Kropp}, {Mine}, {Smy},
  {Sobel}, {Takhistov}, {Weatherly}, {Yankelevich}, {Hill}, {Kim}, {Lim},
  {Park}, {Bodur}, {Scholberg}, {Walter}, {Bernard}, {Drapier}, {Hedri},
  {Giampaolo}, {Gonin}, {Mueller}, {Paganini}, {Quilain}, {Santos}, {Ishizuka},
  {Nakamura}, {Jang}, {Learned}, {Anthony}, {Sztuc}, {Uchida}, {Martin},
  {Scott}, {Berardi}, {Catanesi}, {Radicioni}, {Calabria}, {Machado}, {De
  Rosa}, {Collazuol}, {Iacob}, {Lamoureux}, {Ospina}, {Mattiazzi}, {Ludovici},
  {Nishimura}, {Maewaka}, {Cao}, {Friend}, {Hasegawa}, {Ishida}, {Kobayashi},
  {Jakkapu}, {Matsubara}, {Nakadaira}, {Nakamura}, {Oyama}, {Sakashita},
  {Sekiguchi}, {Tsukamoto}, {Nakano}, {Shiozawa}, {Suzuki}, {Takeuchi},
  {Yamamoto}, {Kotsar}, {Ozaki}, {Ali}, {Ashida}, {Feng}, {Hirota}, {Ichikawa},
  {Kikawa}, {Mori}, {Nakaya}, {Wendell}, {Yasutome}, {Fernandez}, {McCauley},
  {Mehta}, {Tsui}, {Fukuda}, {Itow}, {Menjo}, {Niwa}, {Sato}, {Tsukada},
  {Mijakowski}, {Lagoda}, {Lakshmi}, {Zalipska}, {Jung}, {Vilela}, {Wilking},
  {Yanagisawa}, {Jiang}, {Hagiwara}, {Harada}, {Horai}, {Ishino}, {Ito},
  {Koshio}, {Ma}, {Piplani}, {Sakai}, {Kitagawa}, {Barr}, {Barrow}, {Cook},
  {Goldsack}, {Samani}, {Wark}, {Nova}, {Boschi}, {Di Lodovico}, {Taani},
  {Zsoldos}, {Gao}, {Migenda}, {Yang}, {Jenkins}, {Malek}, {McElwee}, {Stone},
  {Thiesse}, {Thompson}, {Okazawa}, {Nakamura}, {Kim}, {Yu}, {Seo},
  {Nishijima}, {Koshiba}, {Iwamoto}, {Ogawa}, {Yokoyama}, {Martens}, {Vagins},
  {Nakagiri}, {Kuze}, {Izumiyama}, {Yoshida}, {Inomoto}, {Ishitsuka},
  {Matsumoto}, {Ohta}, {Shinoki}, \& {Suganuma}}]{2024PhRvD.110c2003L}
{Locke}, S., {Coffani}, A., {Abe}, K., {et~al.} 2024, \prd, 110, 032003,
  \dodoi{10.1103/PhysRevD.110.032003}

\bibitem[{{Mori} {et~al.}(2021){Mori}, {Suwa}, {Nakazato}, {Sumiyoshi},
  {Harada}, {Harada}, {Koshio}, \& {Wendell}}]{2021PTEP.2021b3E01M}
{Mori}, M., {Suwa}, Y., {Nakazato}, K., {et~al.} 2021, Prog. Theor. Exp. Phys.,
  2021, 023E01, \dodoi{10.1093/ptep/ptaa185}

\bibitem[{Mori {et~al.}(2022)Mori, Abe, Hayato, Hiraide, Ieki, Ikeda, Imaizumi,
  Kameda, Kanemura, Kaneshima, Kashiwagi, Kataoka, Miki, Mine, Miura, Moriyama,
  Nagao, Nakahata, Nakano, Nakayama, Noguchi, Okada, Okamoto, Orii, Sato,
  Sekiya, Shiba, Shimizu, Shiozawa, Sonoda, Suzuki, Takeda, Takemoto, Takenaka,
  Tanaka, Tomiya, Watanabe, Yano, Yoshida, Han, Kajita, Okumura, Tashiro, Wang,
  Xia, Megias, Bravo-Bergu{\~{n}}o, Fernandez, Labarga, Ospina, Zaldivar,
  Zsoldos, Pointon, Blaszczyk, Kearns, Raaf, Stone, Wan, Wester, Bian,
  Griskevich, Kropp, Locke, Smy, Sobel, Takhistov, A., Hill, Kim, Lim, Park,
  Bodur, Scholberg, Walter, Bernard, Coffani, Drapier, Hedri, Giampaolo,
  Mueller, Paganini, Quilain, Santos, Ishizuka, Nakamura, Jang, Learned,
  Anthony, Martin, Scott, Sztuc, Uchida, Berardi, Catanesi, Radicioni,
  Calabria, Machado, Rosa, Collazuol, Iacob, Lamoureux, Mattiazzi, Ludovici,
  Gonin, Pronost, Maekawa, Nishimura, Fujisawa, Friend, Hasegawa, Ishida,
  Kobayashi, Jakkapu, Matsubara, Nakadaira, Nakamura, Oyama, Sakashita,
  Sekiguchi, Tsukamoto, Ozaki, Shiozawa, Suzuki, Takeuchi, Yamamoto, Kotsar,
  Ashida, Bronner, Feng, Hirota, Kikawa, Nakaya, Wendell, Yasutome, McCauley,
  Mehta, Tsui, Fukuda, Itow, Menjo, Ninomiya, Niwa, Tsukada, Lagoda, Lakshmi,
  Mijakowski, Zalipska, Mandal, Prabhu, Jiang, Jung, Vilela, Wilking,
  Yanagisawa, Jia, Hagiwara, Harada, Horai, Ishino, Ito, Kitagawa, Koshio, Ma,
  Nakanishi, Piplani, Sakai, Barr, Barrow, Cook, Samani, Wark, Nova, Boschi,
  Gao, Goldsack, Katori, Lodovico, Migenda, Taani, Zsoldos, Yang, Jenkins,
  Malek, McElwee, Stone, Thiesse, Thompson, Okazawa, Kim, Seo, Yu, Nishijima,
  Koshiba, Nakagiri, Nakajima, Iwamoto, Taniuchi, Yokoyama, Martens, de~Perio,
  Vagins, Kuze, Izumiyama, Yoshida, Inomoto, Ishitsuka, Ito, Kinoshita,
  Matsumoto, Ohta, Ommura, Shigeta, Shinoki, Suganuma, Yamauchi, Martin,
  Tanaka, Towstego, Akutsu, Gousy-Leblanc, Hartz, Konaka, Prouse, Chen, Xu,
  Zhang, Posiadala-Zezula, Hadley, Nicholson, O'Flaherty, Richards, Ali,
  Jamieson, Walker, Marti, Minamino, Okamoto, Pintaudi, Sasaki, Sano, Suzuki,
  Wada, Cao, ichikawa, Nakamura, Tairafune, \& Choi}]{Mori_2022}
Mori, M., Abe, K., Hayato, Y., {et~al.} 2022, ApJ, 938, 35,
  \dodoi{10.3847/1538-4357/ac8f41}

\bibitem[{{Nagakura} {et~al.}(2021){Nagakura}, {Burrows}, {Vartanyan}, \&
  {Radice}}]{2021MNRAS.500..696N}
{Nagakura}, H., {Burrows}, A., {Vartanyan}, D., \& {Radice}, D. 2021, \mnras,
  500, 696, \dodoi{10.1093/mnras/staa2691}

\bibitem[{Nakanishi(2025)}]{zenodo_nakanishi}
Nakanishi, F. 2025, SKSNSim: Supernova burst and Diffuse Supernova Neutrino
  Background simulator for Water Cherenkov Detectors, 1.2.3a,  Zenodo,
  \dodoi{10.5281/zenodo.16751274}

\bibitem[{{Nakanishi} {et~al.}(2024){Nakanishi}, {Izumiyama}, {Harada}, \&
  {Koshio}}]{2024ApJ...965...91N}
{Nakanishi}, F., {Izumiyama}, S., {Harada}, M., \& {Koshio}, Y. 2024, \apj,
  965, 91, \dodoi{10.3847/1538-4357/ad344e}

\bibitem[{Nakazato(2022)}]{zenodo_nakazato}
Nakazato, K. 2022, Supernova Neutrino Light Curves from Proto-Neutron Star
  Cooling with Various Nuclear Equation of State, 1.0,  Zenodo,
  \dodoi{10.5281/zenodo.5778223}

\bibitem[{Nakazato {et~al.}(2013)Nakazato, Sumiyoshi, Suzuki, Totani, Umeda, \&
  Yamada}]{Nakazato_2013}
Nakazato, K., Sumiyoshi, K., Suzuki, H., {et~al.} 2013, ApJS, 205, 2,
  \dodoi{10.1088/0067-0049/205/1/2}

\bibitem[{{Nakazato} \& {Suzuki}(2019)}]{2019ApJ...878...25N}
{Nakazato}, K., \& {Suzuki}, H. 2019, \apj, 878, 25,
  \dodoi{10.3847/1538-4357/ab1d4b}

\bibitem[{{Nakazato} \& {Suzuki}(2020)}]{2020ApJ...891..156N}
---. 2020, \apj, 891, 156, \dodoi{10.3847/1538-4357/ab7456}

\bibitem[{Nakazato {et~al.}(2018)Nakazato, Suzuki, \& Togashi}]{Nakazato_2018}
Nakazato, K., Suzuki, H., \& Togashi, H. 2018, PhRvC, 97,
  \dodoi{10.1103/physrevc.97.035804}

\bibitem[{{Nakazato} {et~al.}(2022){Nakazato}, {Nakanishi}, {Harada}, {Koshio},
  {Suwa}, {Sumiyoshi}, {Harada}, {Mori}, \& {Wendell}}]{Nakazato_2022}
{Nakazato}, K., {Nakanishi}, F., {Harada}, M., {et~al.} 2022, ApJ, 925, 98,
  \dodoi{10.3847/1538-4357/ac3ae2}

\bibitem[{{O'Connor} \& {Ott}(2013)}]{2013ApJ...762..126O}
{O'Connor}, E., \& {Ott}, C.~D. 2013, \apj, 762, 126,
  \dodoi{10.1088/0004-637X/762/2/126}

\bibitem[{{Sato} \& {Suzuki}(1987)}]{1987PhLB..196..267S}
{Sato}, K., \& {Suzuki}, H. 1987, Phys. Lett. B, 196, 267,
  \dodoi{10.1016/0370-2693(87)90728-3}

\bibitem[{{Scholberg}(2012)}]{2012ARNPS..62...81S}
{Scholberg}, K. 2012, Annu. Rev. Nucl. Part. Sci., 62, 81,
  \dodoi{10.1146/annurev-nucl-102711-095006}

\bibitem[{Shen {et~al.}(1998)Shen, Toki, Oyamatsu, \& Sumiyoshi}]{Shen_1998}
Shen, H., Toki, H., Oyamatsu, K., \& Sumiyoshi, K. 1998, Nucl. Phys. A, 637,
  435, \dodoi{https://doi.org/10.1016/S0375-9474(98)00236-X}

\bibitem[{Shen {et~al.}(2011)Shen, Toki, Oyamatsu, \& Sumiyoshi}]{Shen_2011}
---. 2011, ApJS, 197, 20, \dodoi{10.1088/0067-0049/197/2/20}

\bibitem[{Strumia \& Vissani(2003)}]{Strumia_2003}
Strumia, A., \& Vissani, F. 2003, Phys. Lett. B, 564, 42,
  \dodoi{10.1016/s0370-2693(03)00616-6}

\bibitem[{{Sumiyoshi} {et~al.}(2023){Sumiyoshi}, {Furusawa}, {Nagakura},
  {Harada}, {Togashi}, {Nakazato}, \& {Suzuki}}]{2023PTEP.2023a3E02S}
{Sumiyoshi}, K., {Furusawa}, S., {Nagakura}, H., {et~al.} 2023, Prog. Theor.
  Exp. Phys., 2023, 013E02, \dodoi{10.1093/ptep/ptac167}

\bibitem[{{Sumiyoshi} {et~al.}(2005){Sumiyoshi}, {Yamada}, {Suzuki}, {Shen},
  {Chiba}, \& {Toki}}]{2005ApJ...629..922S}
{Sumiyoshi}, K., {Yamada}, S., {Suzuki}, H., {et~al.} 2005, \apj, 629, 922,
  \dodoi{10.1086/431788}

\bibitem[{{Suwa}(2014)}]{2014PASJ...66L...1S}
{Suwa}, Y. 2014, \pasj, 66, L1, \dodoi{10.1093/pasj/pst030}

\bibitem[{{Suwa} {et~al.}(2019){Suwa}, {Sumiyoshi}, {Nakazato}, {Takahira},
  {Koshio}, {Mori}, \& {Wendell}}]{Suwa_2019}
{Suwa}, Y., {Sumiyoshi}, K., {Nakazato}, K., {et~al.} 2019, ApJ, 881, 139,
  \dodoi{10.3847/1538-4357/ab2e05}

\bibitem[{{Suwa} {et~al.}(2022){Suwa}, {Harada}, {Harada}, {Koshio}, {Mori},
  {Nakanishi}, {Nakazato}, {Sumiyoshi}, \& {Wendell}}]{2022ApJ...934...15S}
{Suwa}, Y., {Harada}, A., {Harada}, M., {et~al.} 2022, ApJ, 934, 15,
  \dodoi{10.3847/1538-4357/ac795e}

\bibitem[{{Suwa} {et~al.}(2025){Suwa}, {Harada}, {Mori}, {Nakazato}, {Akaho},
  {Harada}, {Koshio}, {Nakanishi}, {Sumiyoshi}, \&
  {Wendell}}]{2025ApJ...980..117S}
{Suwa}, Y., {Harada}, A., {Mori}, M., {et~al.} 2025, ApJ, 980, 117,
  \dodoi{10.3847/1538-4357/adabe2}

\bibitem[{{Suzuki}(1994)}]{1994pan..conf..763S}
{Suzuki}, H. 1994, in Physics and Astrophysics of Neutrinos, XIII, ed.
  M.~{Fukugita} \& A.~{Suzuki}, 420

\bibitem[{{Takiwaki} \& {Kotake}(2018)}]{2018MNRAS.475L..91T}
{Takiwaki}, T., \& {Kotake}, K. 2018, \mnras, 475, L91,
  \dodoi{10.1093/mnrasl/sly008}

\bibitem[{{Thompson} {et~al.}(2003){Thompson}, {Burrows}, \&
  {Pinto}}]{2003ApJ...592..434T}
{Thompson}, T.~A., {Burrows}, A., \& {Pinto}, P.~A. 2003, \apj, 592, 434,
  \dodoi{10.1086/375701}

\bibitem[{{Togashi} {et~al.}(2017){Togashi}, {Nakazato}, {Takehara},
  {Yamamuro}, {Suzuki}, \& {Takano}}]{2017NuPhA.961...78T}
{Togashi}, H., {Nakazato}, K., {Takehara}, Y., {et~al.} 2017, \nphysa, 961, 78,
  \dodoi{10.1016/j.nuclphysa.2017.02.010}

\bibitem[{{Togashi} \& {Takano}(2013)}]{2013NuPhA.902...53T}
{Togashi}, H., \& {Takano}, M. 2013, \nphysa, 902, 53,
  \dodoi{10.1016/j.nuclphysa.2013.02.014}

\bibitem[{{Woosley} \& {Weaver}(1995)}]{1995ApJS..101..181W}
{Woosley}, S.~E., \& {Weaver}, T.~A. 1995, \apjs, 101, 181,
  \dodoi{10.1086/192237}

\end{thebibliography}
\bibliographystyle{aasjournal}



\end{document}